\documentclass[a4paper,11pt]{article}
\pdfoutput=1 

\usepackage{jcappub} 

\usepackage[T1]{fontenc} 
\usepackage{graphicx}
\usepackage{epsfig}
\usepackage{rotate}
\usepackage{amsmath}
\usepackage{amssymb}
\usepackage{amsfonts}
\usepackage{bm}
\usepackage{enumerate}
\usepackage{afterpage}
\usepackage{xcolor}
\usepackage{natbib, hyperref}

\UseRawInputEncoding

\newcommand{\ud}{\mathrm{d}}
\newcommand{\p}{\partial}
\newcommand{\cH}{\mathcal{H}}
\newcommand{\Q}{\mathcal{Q}}

\def\be{\begin{equation}}
\def\ee{\end{equation}}
\def\bea{\begin{eqnarray}}
\def\eea{\end{eqnarray}}

\title{Detecting the relativistic galaxy bispectrum}

\author{Roy Maartens$^{1,2}$,  Sheean Jolicoeur$^1$, Obinna Umeh$^2$, \\ Eline M. De Weerd$^3$, Chris Clarkson$^{3,1,4}$, Stefano Camera$^{5,6,1}$}
\affiliation{$^1$Department of Physics \& Astronomy, University of the Western Cape, Cape Town 7535, South Africa\\
$^{2}$Institute of Cosmology \& Gravitation, University of Portsmouth, Portsmouth PO1 3FX, UK\\
$^{3}$School of Physics \& Astronomy, Queen Mary University of London, London E1 4NS, UK \\
$^{4}$Department of Mathematics \& Applied Mathematics, University of Cape Town, Cape Town 7701, South Africa\\
$^5$Dipartimento di Fisica, Universit\`a degli Studi di Torino, 10125 Torino, Italy\\
$^6$Istituto Nazionale di Fisica Nucleare, Sezione di Torino, 10125 Torino, Italy}

\abstract{
The Fourier galaxy bispectrum is complex, with the imaginary part arising from leading-order relativistic corrections, due to Doppler, gravitational redshift  and related line-of-sight effects  in redshift space. The detection of the imaginary  part of the bispectrum is potentially a smoking gun signal of relativistic contributions.  We investigate whether next-generation spectroscopic surveys could make such a detection. For a Stage IV spectroscopic $H\alpha$ survey similar to Euclid, we find that  the cumulative signal to noise of this relativistic signature  is $\mathcal{O}(10)$. 
{Long-mode relativistic effects couple to  short-mode Newtonian effects in the galaxy bispectrum, but not in the galaxy power spectrum.
This is  the basis for detectability of relativistic effects in the bispectrum of a single galaxy survey, whereas the power spectrum requires multiple galaxy surveys to detect the corresponding signal.}}

\begin{document}
\maketitle
\flushbottom

\section{INTRODUCTION}

The bispectrum of number count fluctuations in redshift space will become an increasingly important complement to the power spectrum in the extraction of cosmological information from galaxy surveys, 
{in the measurement of {clustering} bias parameters and in the breaking of degeneracies between the clustering amplitude and growth rate.}
Analysis of the Fourier galaxy bispectrum is already well advanced for existing survey data (e.g \cite{Gil-Marin:2016wya,Sugiyama:2018yzo}) and for mock data of future surveys (e.g. \cite{Karagiannis:2018jdt,Yankelevich:2018uaz,Oddo:2019run,Sugiyama:2019ike}).  

Here we highlight a feature of the tree-level Fourier galaxy bispectrum which follows from the  leading-order relativistic contribution -- due to Doppler, gravitational redshift  and related line-of-sight effects -- that is omitted in the standard Newtonian analysis. {These effects generate an imaginary part of the galaxy bispectrum, which can be understood as follows (see also \cite{McDonald:2009ud,Clarkson:2018dwn,Jeong:2019igb} for a more general discussion).  
The Doppler-type contributions to the galaxy density contrast involve one or three derivatives  of scalars along the fixed line of sight $\bm n$ [see \eqref{dg1}, \eqref{dg2} below]. In Fourier space, with the plane-parallel approximation, we have  $\bm{n}\cdot\bm{\nabla} \to {\rm i}\, \bm{n}\cdot\bm{k}$, and this leads to imaginary corrections to the galaxy density contrast, which do not cancel in the bispectrum, unlike in the power spectrum.}
At first order, we have $\delta_g=\delta_{g\rm N}+ \delta_{g\rm D}$, where the Newtonian part $\delta_{g\rm N}$ is real and scales as the linear matter density contrast $\delta$. The 
relativistic Doppler-type part $\delta_{g\rm D}$ scales as ${\rm i}\,(\cH/k) \delta$  (see \cite{McDonald:2009ud, Jeong:2011as, Abramo:2017xnp,Clarkson:2018dwn} and below). At second  order, the relativistic contribution $\delta_{g\rm D}^{(2)}$ scales as ${\rm i}\,(\cH/k) (\delta)^2$ (see \cite{Clarkson:2018dwn} and below).

In the case of  the galaxy {auto}-power spectrum, $P_g\sim \langle |\delta_g|^2\rangle$, the relativistic part is {real and scales as} $(\cH/k)^2P$\,:  therefore we can neglect $P_{g\rm D}$ at leading order. By contrast, for the galaxy bispectrum, $B_g\sim \langle \delta_g\,\delta_g \, \delta^{(2)}_g\rangle$, a coupling of relativistic contributions to short-scale Newtonian terms (which  is absent in $P_g$) produces a $B_{g\rm D}$ that is {imaginary and scales as} ${\rm i}\,(\cH/k)P^2$. 
We therefore expect these relativistic effects to be more accessible in the bispectrum than in the power spectrum, for the case of a single tracer {of the matter distribution}. 

{Although the galaxy bispectrum is statistically isotropic, the plane-parallel approximation in redshift space breaks 3-dimensional isotropy, since a preferred direction is imposed by the observer's fixed line of sight. }

Let us introduce a more explicit analysis, as follows.

\begin{figure}[!h]
\centering
\includegraphics[width=10.0cm]{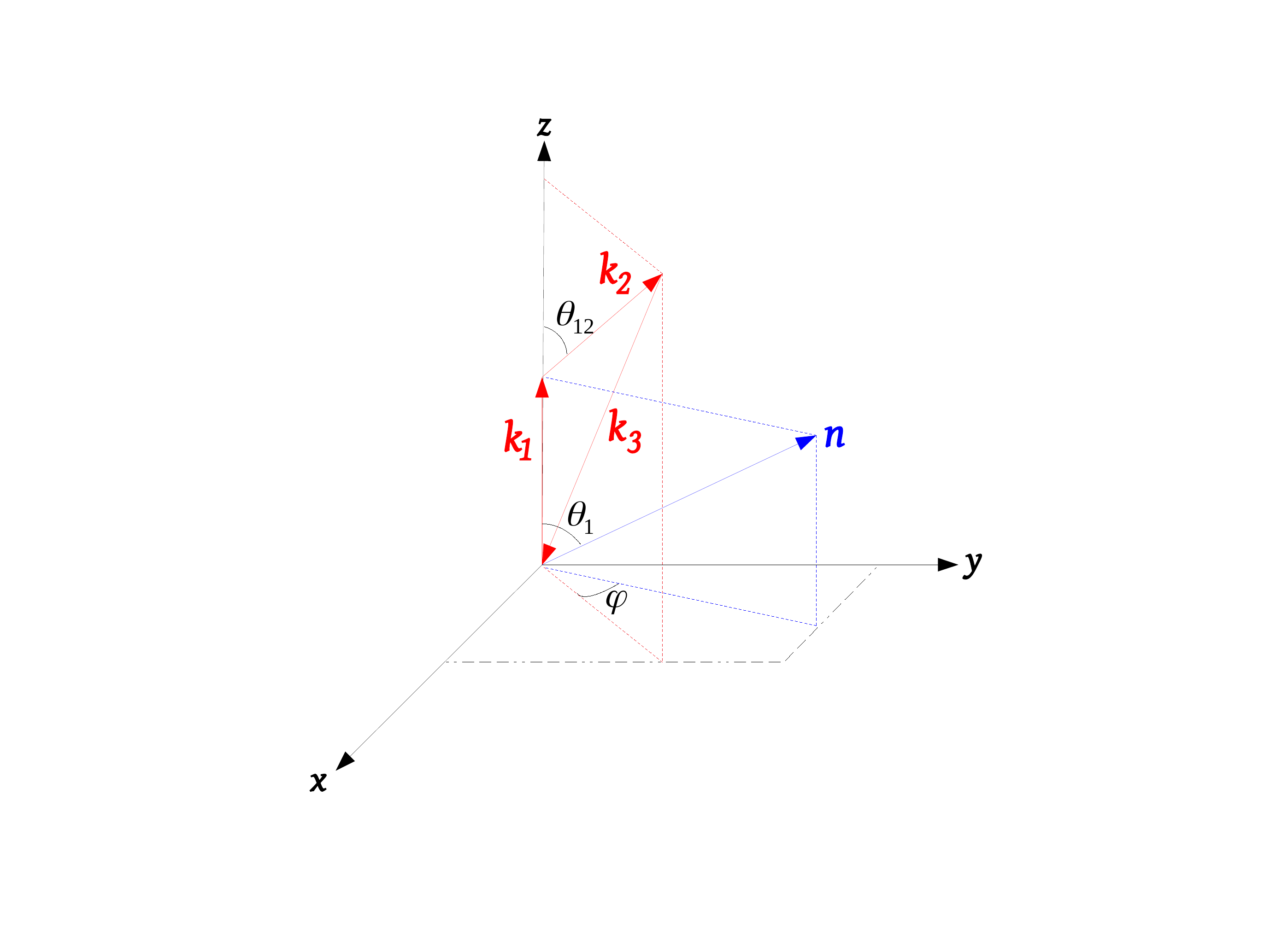}
\vspace*{-1cm}
\caption{Relevant vectors and angles for the Fourier bispectrum.} \label{fig0}
\end{figure}

At tree-level, the Fourier galaxy bispectrum at  a redshift $z$ is given by
\be
{\big\langle \delta_g(z,\bm{k}_{1})\delta_g(z,\bm{k}_{2})\delta^{(2)}_g(z,\bm{k}_{3}) \big\rangle + \text{2 cp}=2 (2\pi)^3 B_{g}(z, \bm{k}_{1}, \bm{k}_{2}, \bm{k}_{3}) \delta^{\rm Dirac}\big(\bm{k}_{1}+ \bm{k}_{2}+ \bm{k}_{3} \big)\,,}
\ee
where cp denotes cyclic permutation and the factor 2 on the right arises from the convention that the total number density contrast is $\delta_g+ \delta^{(2)}_g/2$.  
In terms of the first- and second-order kernels, we have
\begin{equation}
B_{g}(z, \bm{k}_{1}, \bm{k}_{2}, \bm{k}_{3}) = \mathcal{K}^{(1)}(z, \bm{k}_{1})\mathcal{K}^{(1)}(z, \bm{k}_{2})\mathcal{K}^{(2)}(z, \bm{k}_{1}, \bm{k}_{2}, \bm{k}_{3})P(z, k_{1})P(z, k_{2}) + \text{2 cp}\,, \label{e1}
\end{equation}
where $P$ is the linear matter power spectrum. 
The $9-3=6$ degrees of freedom in the triangle condition $\bm{k}_{1}+ \bm{k}_{2}+ \bm{k}_{3}=\bm{0}$ at each $z$ are reduced to 5 by the fixed observer's line of sight direction $\bm{n}$.
The bispectrum can be chosen at each $z$ to be a function  of the 3 magnitudes ${k_a}=\big({k}_{1}, {k}_{2},{k}_{3}\big)$ and 2 angles that define the orientation of the triangle (see Fig. \ref{fig0}):
\be
B_{g}(z, \bm{k}_{a}) =B_{g}(z, {k}_{a},  \mu_1,\varphi) \,.
\ee
Here $\mu_a=\hat{\bm{k}}_a\cdot\bm{n}=\cos\theta_a$,  and $\varphi$ is the angle between the triangle plane and the $(\bm{n},\bm{k}_1)$-plane. The three angles $\theta_{ab}= \cos^{-1}\big(\hat{\bm{k}}_{a} \cdot \hat{\bm{k}}_b\big)$, are determined by $k_a$; then $\mu_2=\mu_1\cos\theta_{12}+ \sin\theta_1\,\sin\theta_{12}\cos\varphi$ is determined when $\varphi$ is given, and $\mu_3=-(\mu_1k_1+\mu_2k_2)/k_3$.

In the standard Newtonian approximation, $B_g=B_{g{\rm N}}$, the kernels in \eqref{e1} contain the galaxy bias and the redshift-space distortions (RSD) at first and second order \cite{Bernardeau:2001qr, Karagiannis:2018jdt}:
\begin{eqnarray}
\mathcal{K}^{(1)}_{\mathrm{N}}(\bm{k}_{1}) &=& b_{1}+f\mu_{1}^{2}\,,  \label{e15} \\  \label{e16}
{\mathcal{K}^{(2)}_{\mathrm{N}}}(\bm{k}_{1}, \bm{k}_{2},{\bm{k}_3}) &=& b_{1}F_{2}(\bm{k}_{1}, \bm{k}_{2}) + b_{2} + f\mu_{3}^{2}G_{2}(\bm{k}_{1}, \bm{k}_{2}) +{fZ_2}(\bm{k}_{1}, \bm{k}_{2})
+ b_{s^{2}}S_{2}(\bm{k}_{1}, \bm{k}_{2}) , \label{k2n}
\eea
where we dropped the $z$-dependence for brevity. Here $f$ is the linear matter growth rate, $b_1,b_2$ are the linear and second-order clustering biases, and $b_{s^{2}}$ is the tidal bias. The kernel  $F_2$ is for second-order density, $G_2 , Z_2$ are for RSD,  and $S_2$  is the kernel for tidal bias {(see the Appendix)}.

The Doppler-type relativistic corrections to the Newtonian number count contrast in redshift space are  given {at first order by \cite{Bonvin:2011bg}:}
\be
\delta_{g\rm D} =  {A}\,\bm{v}\cdot\bm{n}\,,\label{dg1}
\ee
{where $A(z)$ is given below in \eqref{e25} and the momentum conservation equation has been used to  eliminate the gravitational redshift: $\bm{n}\cdot\bm{\nabla}\Phi\equiv \p_r \Phi = -\bm{v}'\!\cdot\bm{n}-\cH\,\bm{v}\cdot\bm{n} $.} Here $\Phi$ is the gravitational potential, $\bm{v}$ is the peculiar velocity,  $\cH$ is the comoving Hubble parameter, and $r$ is the line-of-sight comoving distance.
{Note that $\bm{v}\cdot\bm{n}=\partial_r V$, where $V$ is the velocity potential ($v_i=\partial_iV$).} 
At second order, and neglecting vector and tensor modes, it is shown in
\cite{Clarkson:2018dwn}  that (see also \cite{DiDio:2018zmk}) 
\bea
\delta^{(2)}_{g{\rm D}}&=&A\, \bm{v}^{(2)}\!\!\cdot\bm{n}+2{C}(\bm{v}\cdot\bm{n})\,\delta +2{{E}\over\cH}(\bm{v}\cdot\bm{n})\,\partial_r(\bm{v}\cdot\bm{n})
+{2\over\cH^2}\big[(\bm{v}\cdot\bm{n}) \,\partial_r^2\Phi-\Phi\, \partial_r^2 (\bm{v}\cdot\bm{n}) \big]
\nonumber\\ \label{dg2}
&&{}
 -{2\over \cH}\,\partial_r (\bm{v}\cdot\bm{v})+2{b_1\over\cH}\,\Phi\, \partial_r\delta \,. 
\eea
The redshift-dependent coefficients $C,E$ are  given below  in \eqref{e26}, \eqref{e27}.

In Fourier space, 
neglecting sub-leading $ \mathcal{O}(\cH^2/k^2)$ terms, we find from \eqref{e1} that
\begin{eqnarray}
 B_{g\mathrm{D}}(\bm{k}_{1},\bm{k}_{2},\bm{k}_{3}) &=&  \bigg\{\bigg[\mathcal{K}^{(1)}_{\mathrm{N}}(\bm{k}_{1})\mathcal{K}^{(1)}_{\mathrm{D}}(\bm{k}_{2}) + \mathcal{K}^{(1)}_{\mathrm{D}}(\bm{k}_{1})\mathcal{K}^{(1)}_{\mathrm{N}}(\bm{k}_{2})\bigg]\mathcal{K}^{(2)}_{\mathrm{N}}(\bm{k}_{1},\bm{k}_{2},\bm{k}_{3}) 
\nonumber \\&&{} \quad 
+\mathcal{K}^{(1)}_{\mathrm{N}}(\bm{k}_{1})\mathcal{K}^{(1)}_{\mathrm{N}}(\bm{k}_{2})\mathcal{K}^{(2)}_{\mathrm{D}}(\bm{k}_{1},\bm{k}_{2},\bm{k}_{3})\bigg\}P(k_{1})P(k_{2})+\text{2 cp}. \label{e21}
\end{eqnarray}
The relativistic kernels follow from \eqref{dg1} and \eqref{dg2}; they are given in \cite{Clarkson:2018dwn} as
\begin{eqnarray}
\mathcal{K}^{(1)}_{\mathrm{D}}(\bm{k}_{1}) &=& \mathrm{i}\,\cH f A\,\frac{\mu_{1}}{k_{1}}\,, \label{e23} \\
\mathcal{K}^{(2)}_{\mathrm{D}}(\bm{k}_{1},\bm{k}_{2},\bm{k}_{3}) &=& \mathrm{i}\,\cH f \bigg[
A\,\frac{\mu_{3}}{k_{3}}G_{2}(\bm{k}_{1},\bm{k}_{2})
+C\left(\frac{\mu_{1}}{k_{1}} + \frac{\mu_{2}}{k_{2}}\right)
 +\left(\frac{3}{2}\Omega_{m}-fE\right)\mu_{1}\mu_{2}\left(\frac{\mu_{1}}{k_{2}}+\frac{\mu_{2}}{k_{1}}\right)
\nonumber \\
&&\hspace*{-1.5cm}  
{} -\frac{3}{2}\Omega_{m}\left(\mu_{1}^{3}\frac{k_{1}}{k_{2}^{2}} + \mu_{2}^{3}\frac{k_{2}}{k_{1}^{2}}\right)
+2f\,  {\hat{\bm{k}}_{1} \cdot \hat{\bm{k}}_2}\left(\frac{\mu_{1}}{k_{1}} + \frac{\mu_{2}}{k_{2}}\right) 
 -\frac{3\Omega_{m}b_1}{2f}\left(\mu_{1}\frac{k_{1}}{k_{2}^{2}} + \mu_{2}\frac{k_{2}}{k_{1}^{2}}\right)\!  \bigg]\! .~~~ \label{e24}
\end{eqnarray}
It is clear from \eqref{e21}--\eqref{e24} and from the general expressions given in {\cite{Umeh:2016nuh,Jolicoeur:2017nyt}}, that Doppler-type relativistic effects generate an imaginary correction to the Newtonian bispectrum: 
\be
{{\rm Re}\, B_g =B_{g\mathrm{N}} + \mathcal{O}(\cH^2/k^2)\,,~~ {\rm i}\, {\rm Im}\, B_g = B_{g\mathrm{D}}+ \mathcal{O}(\cH^3/k^3)}\,. \label{bgnd}
\ee

The  coefficients in \eqref{e23} and \eqref{e24} are  \cite{Clarkson:2018dwn}
\begin{eqnarray}
A &=& b_{e} - 2\mathcal{Q} + \frac{2(\mathcal{Q}-1)}{{r} \cH}
 - \frac{\cH'}{\cH^{2}} \label{e25} \,, \\
C &=& b_{1}\big(A+f) + \frac{b_1'}{\cH} + 2\bigg(1-\frac{1}{{r} \cH}\bigg){\frac{\partial b_1}{\partial \ln{L}}\bigg|_{\rm c}} \label{e26}\,, \\
E &=& 4-2A-\frac{3}{2}\Omega_{m} \label{e27} \,,
\end{eqnarray}
where a prime is a conformal time derivative, $\Omega_m=\Omega_{m0}(1+z)H_0^2/\cH^2$,  $L$ is the  luminosity, and $\,|_{\rm c}$ denotes evaluation at the flux cut. 

In addition to the clustering bias  $b_1$, the relativistic bispectrum is sensitive to  the evolution bias and magnification bias, which  are defined as \cite{Alonso:2015uua}
\bea \label{bq}
b_e=  -{\partial \ln n_g \over \partial\ln (1+z)}\,,~~~{ {\cal Q} =-{\partial \ln n_g \over \partial\ln L}\bigg|_{\rm c}}\,.
\eea
Here and below,  $n_{g}$ is the {\em comoving} galaxy number density.  (Note that the alternative magnification bias parameter  $s=2{\cal Q}/5$ is often used.)

{It is interesting to note that the magnification bias ${\cal Q}$ enters the relativistic bispectrum, even though we have not  included the effect of the integrated lensing magnification $\kappa$. The reason for this apparent inconsistency is that there is a (non-integrated) Doppler correction to $\kappa$ at leading order \cite{Bonvin:2008ni,Bolejko:2012uj}.}

\section{SIGNAL TO NOISE}

The signal-to-noise ratio (SNR) for the bispectrum  at redshift $z$ is given in the Gaussian approximation of uncorrelated triangles by \cite{Scoccimarro:2003wn}
\begin{equation}
\bigg[\frac{S}{N}(z)\bigg]^{2} = 
\sum_{k_a,\,\mu_{1},\,\varphi}\,{1\over {\rm Var} [{B_{g}}(z, k_a,\mu_{1},\varphi)]}
\,B_{g}(z, k_{a},  \mu_{1},\varphi)\,B^*_{g}(z, k_a, \mu_{1},\varphi)\,,\label{e3}
\end{equation} 
where we have introduced the complex conjugate $B^*_{g}$ since the bispectrum  has an imaginary correction. {Here ${\rm Var} [{B_{g}}]$ is the variance  of the bispectrum estimator
 \cite{Chan:2016ehg}:
\be \label{hatb}
\hat{B}_g(z,\bm{k}_a) = {k_{\rm f}^3 \over V_{123}}\int_{\bm{k}_a}\ud^3\bm{q}_1\, \ud^3\bm{q}_2 \,\ud^3\bm{q}_3\,\delta^{{\rm Dirac}}(\bm{q}_1+\bm{q}_2+\bm{q}_3)\, \delta_g(z,\bm{q}_1) \delta_g(z,\bm{q}_2) \delta_g(z,\bm{q}_3) \,,
\ee 
where integration is over the shells $k_a-\Delta k/2\leq q_a \leq k_a+\Delta k/2$ and  the shell volume is
$V_{123}=\int_{\bm{k}_a}\ud^3\bm{q}_1\, \ud^3\bm{q}_2 \,\ud^3\bm{q}_3\,\delta^{{\rm Dirac}}(\bm{q}_1+\bm{q}_2+\bm{q}_3)$.}

In the Newtonian approximation, the Gaussian variance can be given as \cite{Scoccimarro:2003wn, Karagiannis:2018jdt}
\be
{{\rm Var} [{B_{g}}(z, k_a,\mu_{1},\varphi)]}
=s_B\,{\pi k_{\rm f}(z)^3 \over k_1k_2k_3 (\Delta k)^3}\,{N_{\mu_1}N_\varphi \over \Delta \mu_1 \Delta \varphi} \, \tilde{P}_{g{\rm N}}(z,k_{1},\mu_{1}) \tilde{P}_{g{\rm N}}(z,k_{2},\mu_{2})\tilde{P}_{g{\rm N}}(z,k_{3},\mu_{3})\,,
\label{e4}
\ee  
where 
\begin{equation}
\tilde{P}_{g{\rm N}}(z, k_{a}, \mu_{a}) = P_{g{\rm N}}(z, k_{a}, \mu_{a}) + \frac{1}{n_{g}(z)}\,, \label{e5}
\end{equation}
and ${P}_{g{\rm N}}=(b_1+f\mu_a^2)^2P$ is the linear galaxy power spectrum.
In \eqref{e4}, $s_{B}$ is 6, 2, 1 respectively for equilateral, isosceles and non-isosceles triangles, and $N_{\mu_1},N_\varphi$ are the ranges for $\mu_1,  \varphi $ (which are sometimes reduced from their full values of 2 and $2\pi$ using symmetry arguments).
The fundamental mode is determined by the comoving survey volume of the redshift bin centred at $z$, i.e.,  $k_{\mathrm{f}}(z) = {2\pi}{V(z)^{-1/3}}$, where $V(z)=4\pi  f_{\rm sky}[r(z+\Delta z/2)^3 - r(z-\Delta z/2)^3]$.

For a survey with redshift bin centres ranging from $z_{\rm min}$ to $z_{\rm max}$,  the cumulative SNR  is
\bea \label{csnr}
\frac{S}{N}\big(\leq z\big) = \left\{\sum_{z'=z_{\rm min}}^z \left[\frac{S}{N}(z')\right]^{2} \right\}^{1/2} \,,
\eea
and then the total SNR is $S/N(\leq z_{\rm max})$.

~\\
\subsection{{Relativistic contribution to the variance}}
~\\
{For the full bispectrum, including the relativistic part, \eqref{hatb}}
leads to a variance of the form 
\be \label{varb}
{\rm Var} [{B_{g}}(z,\bm{k}_a)] \propto \tilde{P}_{g}(z,k_{1},\mu_{1}) \tilde{P}_{g}(z,k_{2},\mu_{2})\tilde{P}_{g}(z,k_{3},\mu_{3}) \,.
\ee
In the Newtonian approximation, this gives \eqref{e4}. By \eqref{e23}, the galaxy number density contrast has an imaginary relativistic correction, $\delta_g=\delta_{g{\rm N}}+\delta_{g{\rm D}}$. However, since $P_g \sim\langle \delta_g(\bm{k}) \delta_g(\bm{-k})\rangle=\langle |\delta_g(\bm{k})|^2\rangle$, the galaxy power spectrum  is given by \cite{McDonald:2009ud,Abramo:2017xnp,Clarkson:2018dwn}
\be \label{pgnd}
P_g= P_{g{\rm N}} + P_{g{\rm D}}=  P_{g{\rm N}}+ \mathcal{O}(\cH^2/k^2) \,.
\ee
 It follows from \eqref{varb} and \eqref{pgnd} that  at leading order, the relativistic contribution to the variance can be neglected:
\be
{{\rm Var} [{B_{g}}] = {\rm Var} [{B_{g{\rm N}}}] + \mathcal{O}(\cH^2/k^2)\,.}
\ee
 Therefore the SNR for the Newtonian and relativistic parts of the bispectrum are
 \bea
\bigg(\frac{S}{N}\bigg)^{2}_{\rm N} &=& 
\sum_{k_a,\,\mu_{1},\,\varphi}\,{B_{g{\rm N}}\,B_{g{\rm N}} \over {{\rm Var} [{B_{g{\rm N}}}]}}\,,\label{e3a}\\
\bigg(\frac{S}{N}\bigg)^{2}_{\rm D} &=& 
\sum_{k_a,\,\mu_{1},\,\varphi}\,{B_{g{\rm D}}\,{B^*_{g{\rm D}}} \over {{\rm Var} [{B_{g{\rm N}}}]}}\,.\label{e3b}  
\eea

~\\
\subsection{{Nonlinear effects}} 
~\\

{In order to avoid nonlinear effects of matter clustering, the maximum  $k$ is chosen as a scale where perturbation theory for the matter density contrast begins to break down. It is known that the matter bispectrum is more sensitive to nonlinearity than the matter power spectrum: at $z\sim 0$ nonlinearity sets in at $k\sim 0.1h/\,$Mpc for the matter bispectrum, as opposed to $k\sim 0.2h/\,$Mpc for the matter power spectrum. To  account for the growth of $k_{\rm max}$ with redshift, we use the redshift-dependence proposed in \cite{Smith:2002dz} for the power spectrum, but with half the amplitude at $z=0$:}
\be\label{kmax}
k_{\mathrm{max}}(z) = {0.1h}\big(1+z\big)^{2/(2+n_{s})}\,.
\ee

The  cut-off $k\leq k_{\rm max}(z)$  avoids a breakdown of perturbative accuracy in the matter correlations, but nonlinearities in the galaxy correlations due to RSD can affect longer wavelength modes. The effect of RSD on these scales is to damp the power -- the  `FoG' effect. In order to take account of this, we follow \cite{Karagiannis:2018jdt,Yankelevich:2018uaz} and use the simple model of FoG damping,
\bea
&& P_g \to  D_P\, P_g\,,\quad~ D_{P}(z, \bm{k}) = \exp\Big\{-{1\over2}{\big[k\mu\,\sigma(z)\big]^{2}}\Big\}\,, \\
&&B_g \to  D_B\, B_g\,,  \quad D_{B}(z, \bm{k}_{1},\bm{k}_{2},\bm{k}_{3}) = \exp\Big\{-{1\over2}{\left[k_{1}^{2}\mu_{1}^{2}+k_{2}^{2}\mu_{2}^{2}+k_{3}^{2}\mu_{3}^{2}\right]\sigma(z)^{2}}\Big\}\,, \label{e13}
\eea
where $\sigma$ is the linear velocity dispersion.

On sufficiently large scales the non-Gaussian contribution to the bispectrum covariance can be  approximated by including corrections to the power spectra appearing in the bispectrum variance \eqref{e4}. This is shown by  \cite{Chan:2016ehg}  (see also \cite{Karagiannis:2018jdt}), using  the approximation:
\bea
{\rm Var} [{B_{g}}] &\to & {\rm Var} [{B_{g}}]+ \delta {\rm Var} [{B_{g}}] \,, \\
\delta {\rm Var} [{B_{g}}] &=& {s_B\pi k_{\rm f}^3N_{\mu_1}N_\varphi \over k_1k_2k_3 (\Delta k)^3\Delta \mu_1 \Delta \varphi}\, \Big\{\tilde{P}_{g{\rm N}}(1) \tilde{P}_{g{\rm N}}(2)\Big[ \tilde{P}^{\rm NL}_{g{\rm N}}(3)-\tilde{P}_{g{\rm N}}(3)\Big] +2\, \mbox{cp}\Big\}\,.
\label{e4x}
\eea
Here $\tilde{P}_{g{\rm N}}(a)\equiv \tilde{P}_{g{\rm N}}(z,k_a,\mu_a)$ and $\tilde{P}^{\rm NL}_{g{\rm N}}(a) = (b+f\mu_a^2)^2P^{\rm NL}+n_g^{-1}$, where $P^{\rm NL}$ is the nonlinear matter power spectrum, computed with a modified Halofit emulator.

~\\
\subsection{Summations {over triangles}} 
~\\
The counting of triangles $\bm{k}_{1}+\bm{k}_{2}+\bm{k}_{3}=\bm{0}$ that contribute to the signal-to-noise  involves a sum in $k_a$-space and a sum over orientations.

The triangle sides are chosen so that $k_1\geq k_2 \geq k_3$, and must satisfy  {$k_1-k_2-k_3\leq 0$}.
For the summation in $k_a$ we choose the minimum and  the step-length as
\be
k_{\rm min}(z)=k_{\rm f}(z)~~\mbox{and}~~ \Delta k(z)=k_{\mathrm{f}}(z) \,,
\ee
as in \cite{Karagiannis:2018jdt,Yankelevich:2018uaz}. 
Then the $k_a$ sum is defined as \cite{Liguori:2010hx,Oddo:2019run}
\begin{equation}
 {\sum_{k_a} = \sum_{k_{1}=k_{\mathrm{min}}}^{k_{\mathrm{max}}}\,\sum_{k_{2}=k_{\mathrm{min}}}^{k_{1}}\,\sum_{k_{3}=k_{\mathrm{min}}}^{k_{2}}}
\,.\label{e8}
\end{equation} 

The  coordinates $(\mu_1=\cos\theta_{1},\varphi)$ describe all possible orientations of the triangle. 
We follow  \cite{Karagiannis:2018jdt} and choose the ranges $N_{\mu_1}=2, N_{\varphi}=2\pi$.
For a given $\mu_{1}$, a complete rotation in $\varphi$ about $\bm{k}_1$ double counts the  triangle falling onto the fixed $(\bm{n},\bm{k}_1)$-plane at $\varphi=0$ and $\varphi=2\pi$ (see Fig. \ref{fig0}). Similarly, for a given $\varphi$,  the end-points $\theta_1=0$ and $\theta_1=\pi$  correspond to equivalent triangles, with $\bm{k}_a \to - \bm{k}_a$.
This double-counting can be avoided by imposing suitable upper limits:
$-1\leq \mu_1 <1$ and $0\leq \varphi < 2\pi$.
The signal to noise is quite sensitive to the step-lengths $\Delta \mu_{1}, \Delta\varphi$. We find (see the Appendix  for details) that a suitable choice for convergence is
\bea\label{ori}
\Delta \mu_{1}=0.04\,,~~ \Delta\varphi= \pi/25 \,.
\eea

\section{GALAXY SURVEY} 
 
 \begin{figure}[! h]
\centering
\includegraphics[width=5.cm]{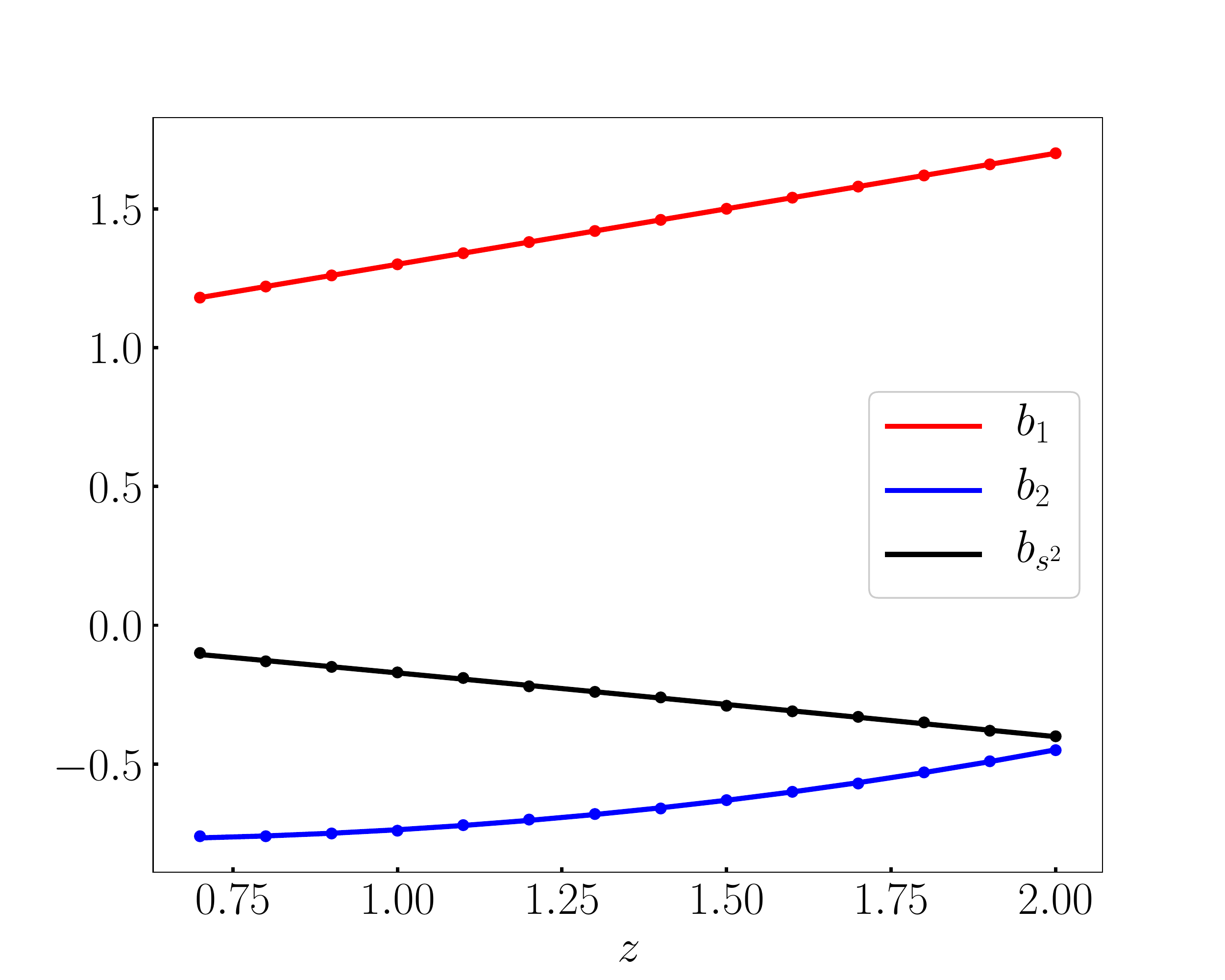}
\includegraphics[width=5.cm]{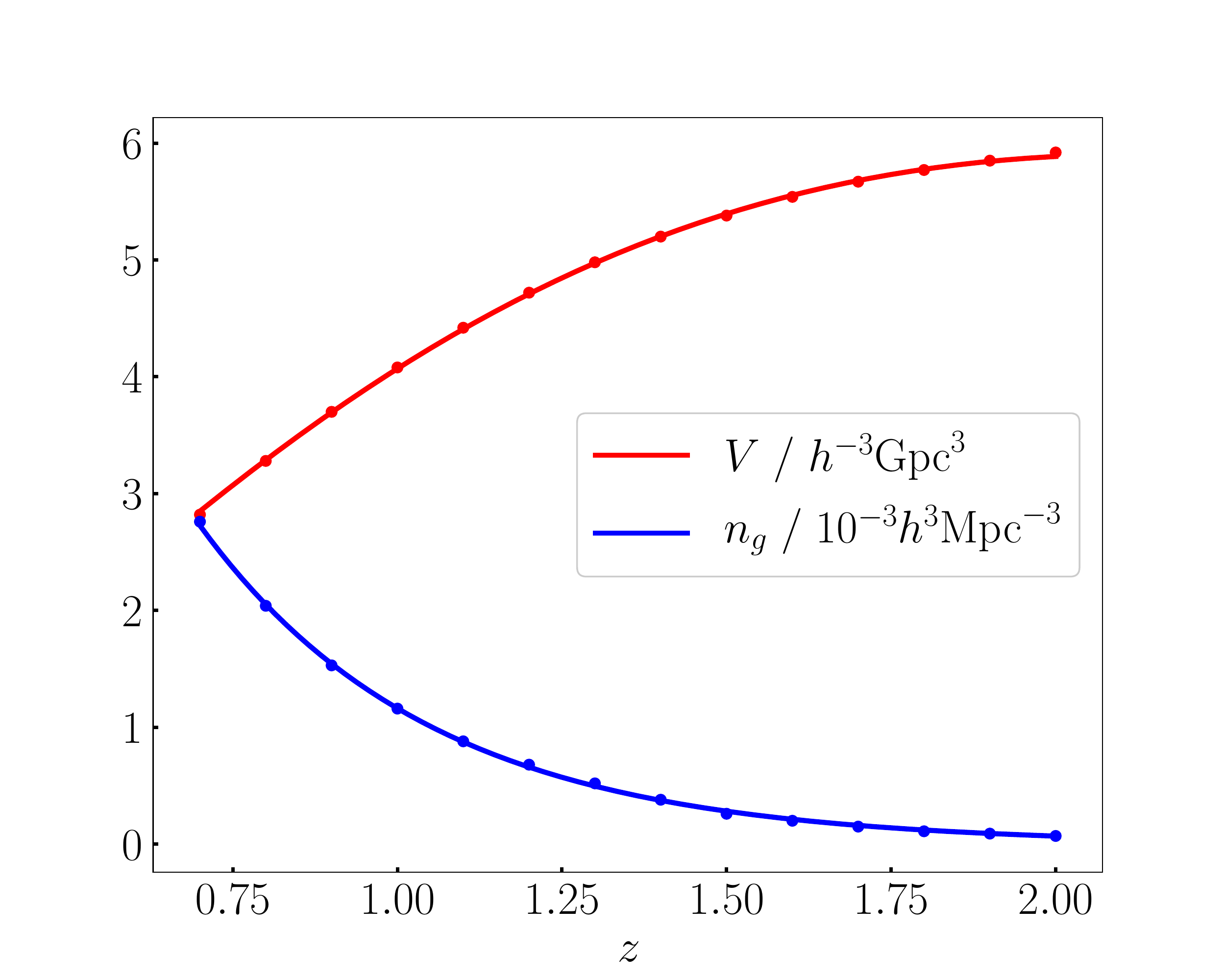} 
\includegraphics[width=5.cm]{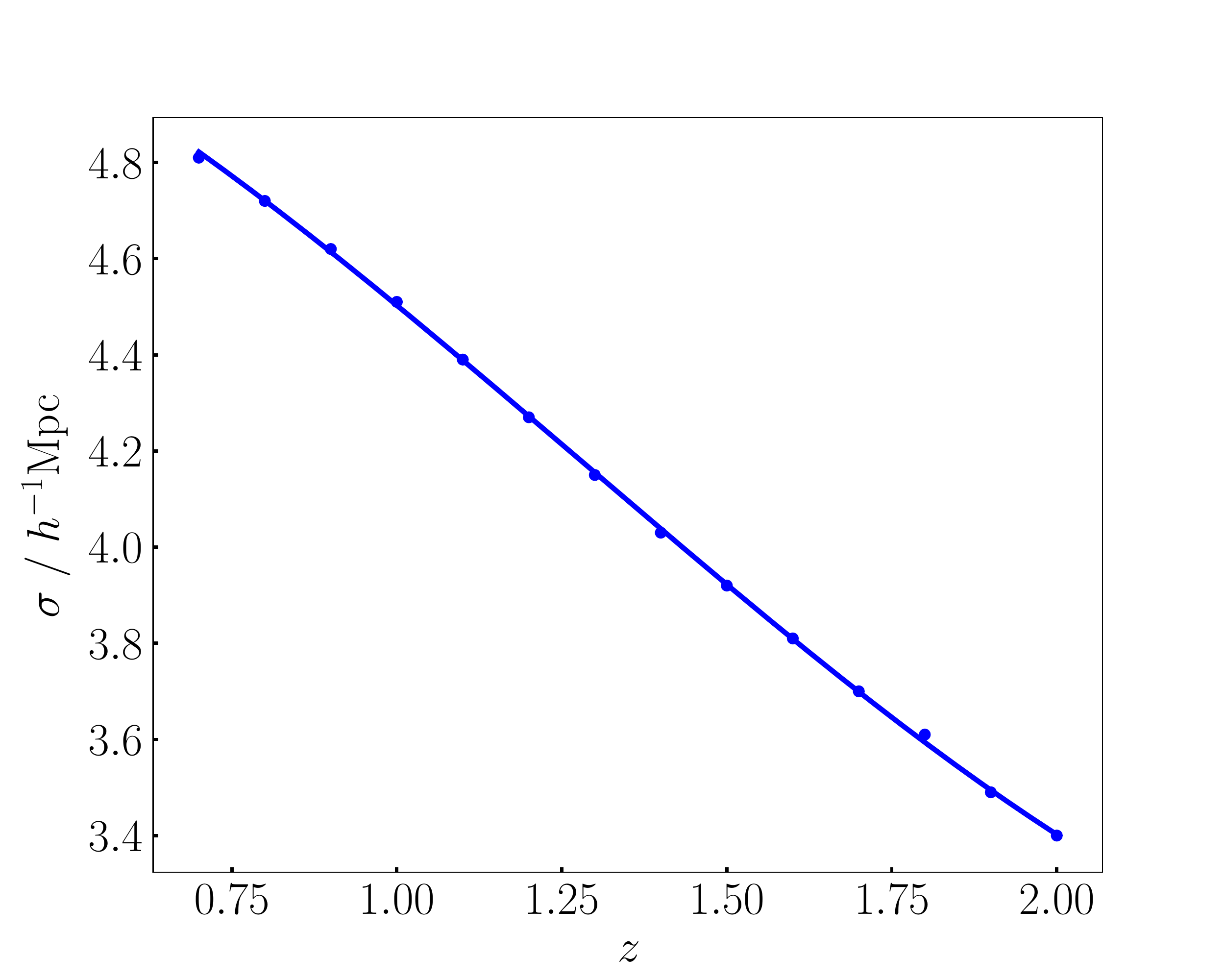}
\caption{Clustering bias parameters   ({\em left}), comoving volume and number density  ({\em middle}) and RSD damping parameter ({\em right}).  Points are the data from Table 1 in \cite{Yankelevich:2018uaz}.}
\label{fig1}
\end{figure}

We consider a Stage IV $H \alpha$ spectroscopic survey, with clustering bias, comoving volume, comoving  number density and RSD damping parameter given by Table 1 in \cite{Yankelevich:2018uaz}, over the redshift range {$0.65\leq z\leq 2.05$}, with  $\Delta z=0.1$ bins. 
We provide fitting formulas for these quantities in the Appendix.
Figure \ref{fig1} shows the values given in \cite{Yankelevich:2018uaz} together with the fitting curves. For the  cosmological parameters, we use Planck 2018  \cite{Aghanim:2018eyx}:
$h = 0.6766, \Omega_{m0} = 0.3111,
 \Omega_{b0}h^{2}= 0.02242,
\Omega_{c0} h^{2} = 0.11933,
n_s = 0.9665,
\sigma_8 = 0.8102, \gamma=0.545,
\Omega_{K0} = 0= \Omega_{\nu 0}$.

We checked that the SNR for the Newtonian bispectrum is consistent with Fig.~4 of \cite{Yankelevich:2018uaz}, when we use their redshift-independent $k_{\mathrm{max}} = 0.15h\;\mathrm{Mpc}^{-1}$, {and when we remove the flattened triangle shapes that are excluded by \cite{Yankelevich:2018uaz}. When we include the flattened shapes, we checked that we recover the total number of triangles given in Table~1 of \cite{Oddo:2019run}.}

~\\
\subsection{Evolution bias and magnification bias} 
~\\
The relativistic bispectrum depends  also on $b_e$  and ${\cal Q}$, as shown in \eqref{e24}--\eqref{e27}.
These parameters do {\em not} appear in the Newtonian approximation, but they are crucial for the relativistic correction, and we need to evaluate them in a physically consistent way. We compute these parameters from the same luminosity function that is  used to generate the number density shown in Fig. \ref{fig1}, i.e., Model~1 in \cite{Pozzetti:2016cch}:
\begin{equation}
\Phi( z,y) = \Phi_{*}(z) y^\alpha\,{\rm e}^{-y}\,, \quad y\equiv {L\over L_{*}}\,.
\label{e12_3}
\end{equation}
We have written $\Phi$ in terms of the redshift $z$ and the normalised dimensionless luminosity $y$, where $L_*=L_{*0}(1+z)^\delta$ and $L_{*0}$ is a characteristic luminosity.   Here
 $\alpha$ is the faint-end slope, 
and $\Phi_{*}$ is a characteristic comoving density of $H\alpha$ emitters, modelled as 
\begin{eqnarray}
{\Phi_{*} \over \Phi_{*0}}= \,
\begin{cases} (1+z)^{\epsilon} &   z \leq z_{\mathrm{b}}\,,\\
(1+z_{\mathrm{b}})^{2\epsilon}(1+z)^{-\epsilon} & z > z_{\mathrm{b}} \,.
\end{cases}
\end{eqnarray}
The best-fit parameters for Model 1 are given by \cite{Pozzetti:2016cch} as 
\be \label{eucp}
\alpha=-1.35\,, ~~\delta=2\,,~~ L_{*0}=10^{41.5}\,\mathrm{erg\,s}^{-1}\,,~~\Phi_{*0}=10^{-2.8}\,\mathrm{Mpc}^{-3}\,,~~\epsilon=1\,,~~z_{\mathrm{b}}=1.3  \,. 
\ee
The flux cut  
 $F_{\mathrm{c}}$
 translates to a luminosity cut:
 \be \label{lc}
 L_{\rm c}(z)=4\pi F_{\rm c} \,d_L(z)^2\,,\quad F_{\mathrm{c}} =3\times10^{-16}\,\mathrm{erg\,cm^{-2}\,s^{-1}} \,,
 \ee
where $d_L$ is the background luminosity distance and the choice of $F_{\rm c}$ follows \cite{Yankelevich:2018uaz}. 
\begin{figure}[!h]
\centering
\includegraphics[width=8.0cm]{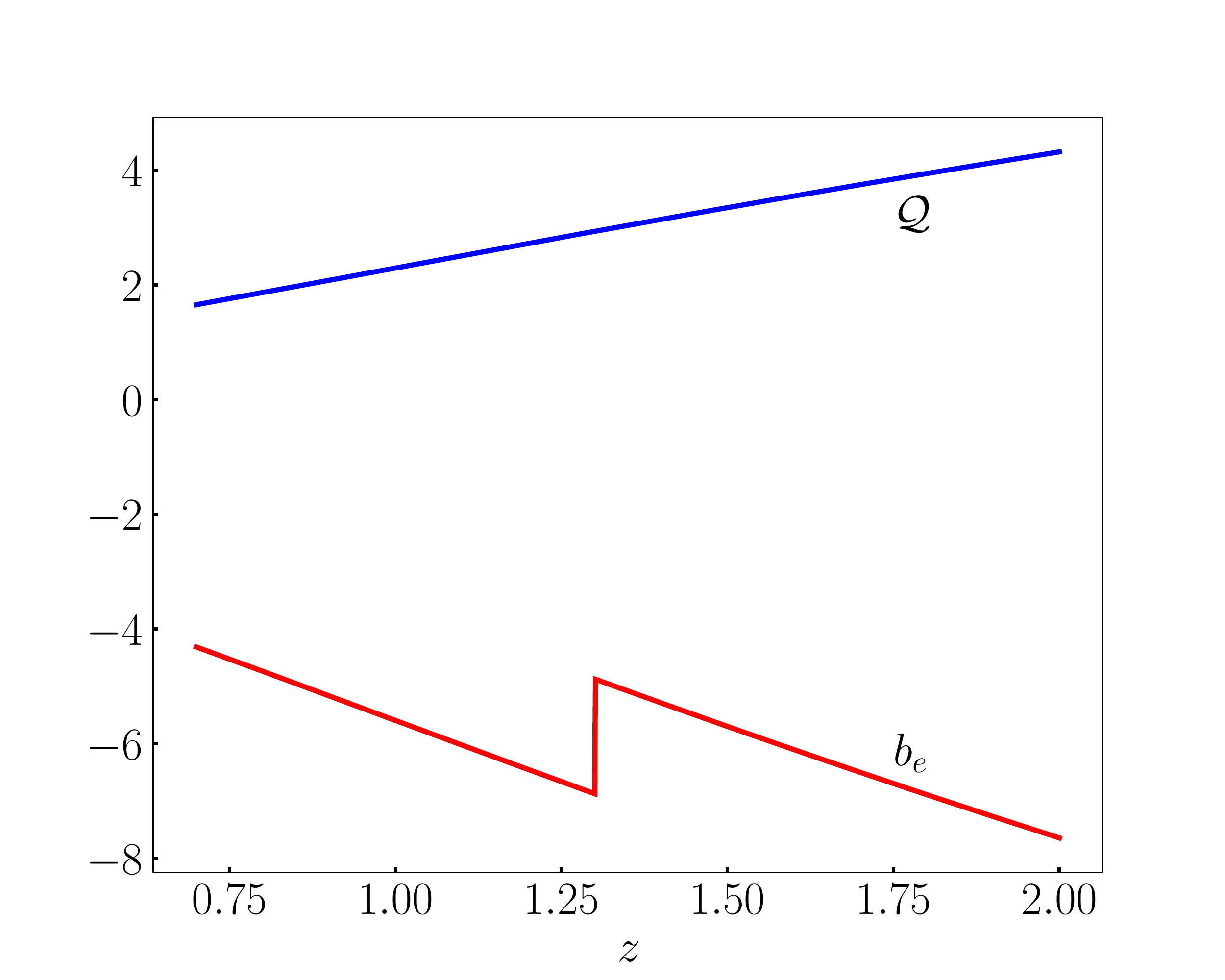}
\caption{Magnification and evolution bias \eqref{q}, \eqref{be}.} 
\label{fig2}
\end{figure}

In order to compute $b_{e}$ and $\Q$, we require the comoving number density
\be \label{ng}
n_g(z)=\int_{y_c(z)}^\infty {\ud  y}\, \Phi(z,y) = \Phi_*(z) \, \Gamma({\alpha+1}, y_c(z)) \,,
\ee
where $\Gamma$ is the upper incomplete Gamma function and
\be  \label{ycz}
y_c(z)={4\pi F_{\rm c} \over L_{*0}}\, {r}(z)^2  =\left[ {{r}(z) \over 2.97\,h\times 10^3~ ({\rm Mpc}/h)}\right]^2 \,.
\ee
Using \eqref{ycz} and \eqref{e12_3}--\eqref{lc},  we confirm that the analytical form \eqref{ng} for $n_g$ recovers the points from Table~1 in \cite{Yankelevich:2018uaz}. 

By \eqref{bq}, the magnification bias follows as
\be\label{q}
{\cal Q}(z) ={\left( y\, {\Phi \over n_g}\right)_{\rm c}}= { y_{\rm c}(z)^{{\alpha+1}}\exp\big[-y_{\rm c}(z) \big] \over \Gamma({\alpha+1}, y_c(z))}\,,
\ee
since {$\partial/\partial \,{\ln L} = \partial/\partial\, {\ln y}$},  and the evolution bias is \footnote{Error in equation (3.8) fixed - results unchanged.} \citep{Maartens:2021dqy}
\be\label{be}
b_e(z) = -{\ud \ln  \Phi_*(z)\over \ud \ln (1+z)}-  { {\ud \ln L_{*} }(z) \over \ud \ln (1+z)}\,{\cal Q}(z)\,.
\ee
Figure \ref{fig2} shows the analytical forms \eqref{q} and \eqref{be}  for  $b_{e}$ and $\Q$.  

Table \ref{tab1} collects the information in Figs. \ref{fig1} and \ref{fig2} to provide an extension of Table  1 in \cite{Yankelevich:2018uaz}.
\begin{table}[!ht] 
\centering 
\caption{Stage IV $H\alpha$ spectroscopic survey parameters.} \label{tab1} 
\vspace*{0.2cm}
\begin{tabular}{|c|c|c|c|c|c|c|c|c|} 
\hline 
~~$z$~~ & ~~$b_{1}$~~ & ~~~~$b_{2}$~~~~ & ~~~~$b_{s^{2}}$~~~~ & ~~~$b_{e}$~~~ & ~~~$\mathcal{Q}$~~~ & $n_{g}$ & $V$ & $\sigma$ \\ 
 & & & & & & $10^{-3}h^{3} \mathrm{Mpc}^{-3}$ & $h^{-3} \mathrm{Gpc}^{3}$ & $h^{-1} \mathrm{Mpc}$\\
\hline\hline 
0.7 & 1.18 & -0.766 & -0.105 & -4.31 & 1.66 & 2.76 & 2.82 & 4.81 \\
0.8 & 1.22 & -0.759 & -0.127 & -4.74 & 1.87 & 2.04 & 3.38 & 4.72 \\
0.9 & 1.26 & -0.749 & -0.149 & -5.17 & 2.08 & 1.53 & 3.70 & 4.62 \\
1.0 & 1.30 & -0.737 & -0.172 & -5.60 & 2.30 & 1.16 & 4.08 & 4.51 \\
1.1 & 1.34 & -0.721 & -0.194 & -6.02 & 2.51 & 0.880 & 4.42 & 4.39 \\
1.2 & 1.38 & -0.703 & -0.217 & -6.45 & 2.72 & 0.680 & 4.72 & 4.27 \\
1.3 & 1.42 & -0.682 & -0.240 & -6.76 & 2.94 & 0.520 & 4.98 & 4.15 \\
1.4 & 1.46 & -0.658 & -0.262 & -5.29 & 3.14 & 0.380 & 5.20 & 4.03 \\
1.5 & 1.50 & -0.631 & -0.285 & -5.70 & 3.35 & 0.260 & 5.38 & 3.92 \\
1.6 & 1.54 & -0.600 & -0.308 & -6.10 & 3.55 & 0.200 & 5.54 & 3.81 \\
1.7 & 1.58 & -0.567 & -0.332 & -6.50 & 3.75 & 0.150 & 5.67 & 3.70 \\
1.8 & 1.62 & -0.531 & -0.355 & -6.89 & 3.94 & 0.110 & 5.77 & 3.61 \\
1.9 & 1.66 & -0.491 & -0.378 & -7.27 & 4.13 & 0.0900 & 5.85 & 3.49 \\
2.0 & 1.70 & -0.449 & -0.401 & -7.64 & 4.32 & 0.0700 & 6.92 & 3.40 \\ \hline
\end{tabular}
\end{table}

Finally, we need to deal with the luminosity derivative of the bias in \eqref{e26}. Simulations by \cite{Pan:2019dxa} indicate that the  clustering bias of $H\alpha$ galaxies does not vary appreciably with luminosity near the fiducial luminosity $L_{*0}$ in \eqref{eucp} and for $z\lesssim 2$ (see their Fig. 8). We therefore take
\be\label{b1l}
{\partial b_1 \over \partial \ln{L}}\bigg|_{\rm c}=0\,, 
\ee
in \eqref{e26}. 

~\\
\subsection{Signal to noise of the relativistic bispectrum}
~\\
We can now evaluate the Doppler-type relativistic part of the bispectrum, \eqref{e21}--\eqref{e27}, using \eqref{q}--\eqref {b1l}. 
Then the SNR is computed using  \eqref{csnr}  and \eqref{e3b} together with \eqref{e4x}.
The results, for SNR in each $z$-bin, $S/N(z)$, and for the cumulative SNR, $S/N(\leq z)$, are shown in Fig.~\ref{fig4}.
Our forecasts indicate that the total SNR, $S/N(\leq z_{\rm max})$, for a Stage IV $H\alpha$  survey could be ${\cal O}(10)$, which  is high enough for a detection in principle.

The relativistic SNR is sensitive in particular to two factors:
\begin{itemize}
\item 
Changes in the nonperturbative scale  $k_{\rm max}(z)$: this sensitivity is due to the coupling of long-wavelength relativistic terms to short-wavelength Newtonian terms. 
We use a conservative and redshift-dependent $k_{\rm max}$, given in \eqref{kmax}. In  Fig.~\ref{kmsnr}
we show the comparison of SNR using \eqref{kmax} and using the redshift-independent $k_{\rm max}=0.15h/\,$Mpc. The redshift-independent model does not incorporate  the increase in the nonperturbative scale with growing $z$, and therefore produces a lower SNR; however, the difference is not large.
\begin{figure}[! h]
\centering
\includegraphics[width=7.5cm]{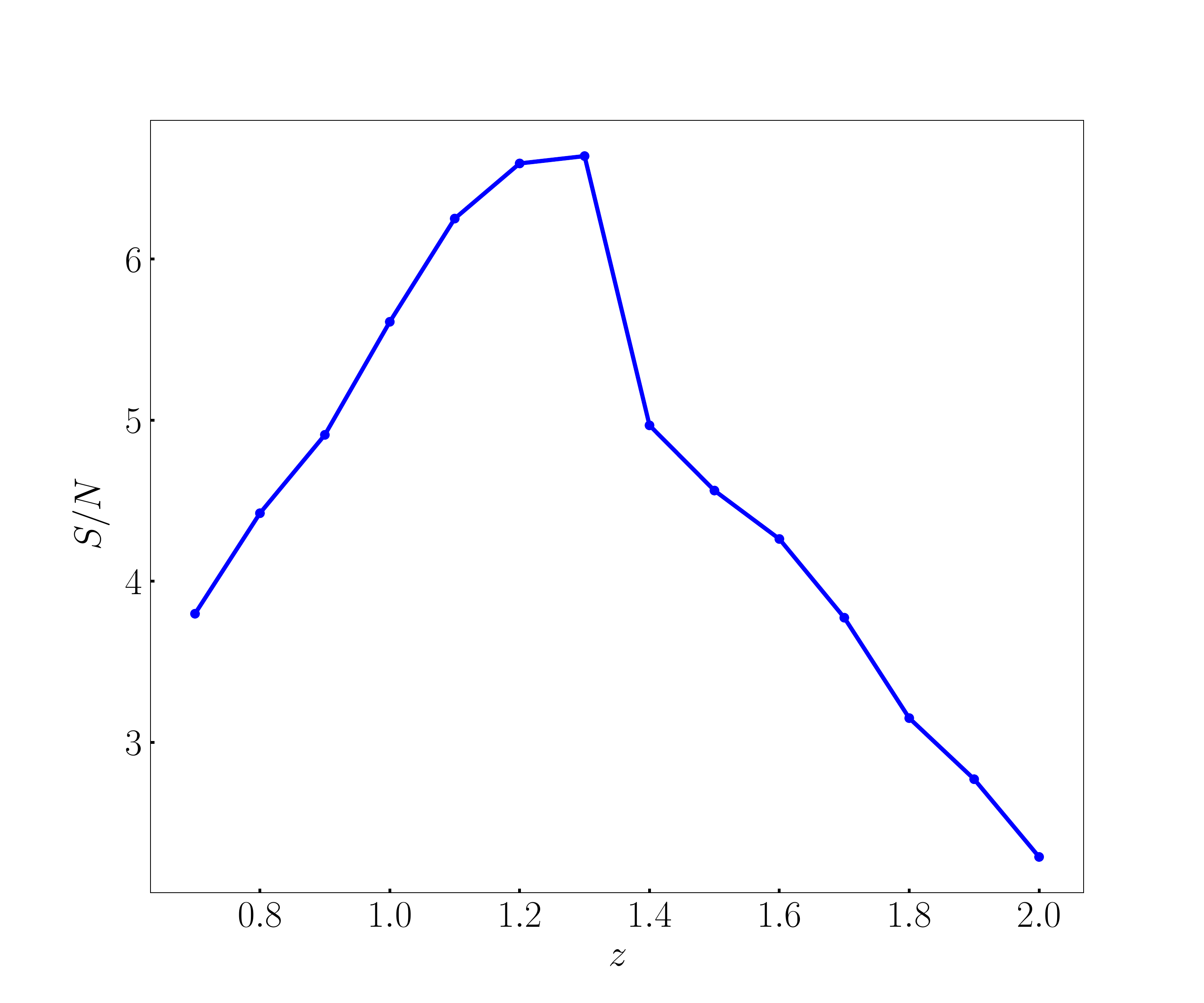}
\includegraphics[width=7.5cm]{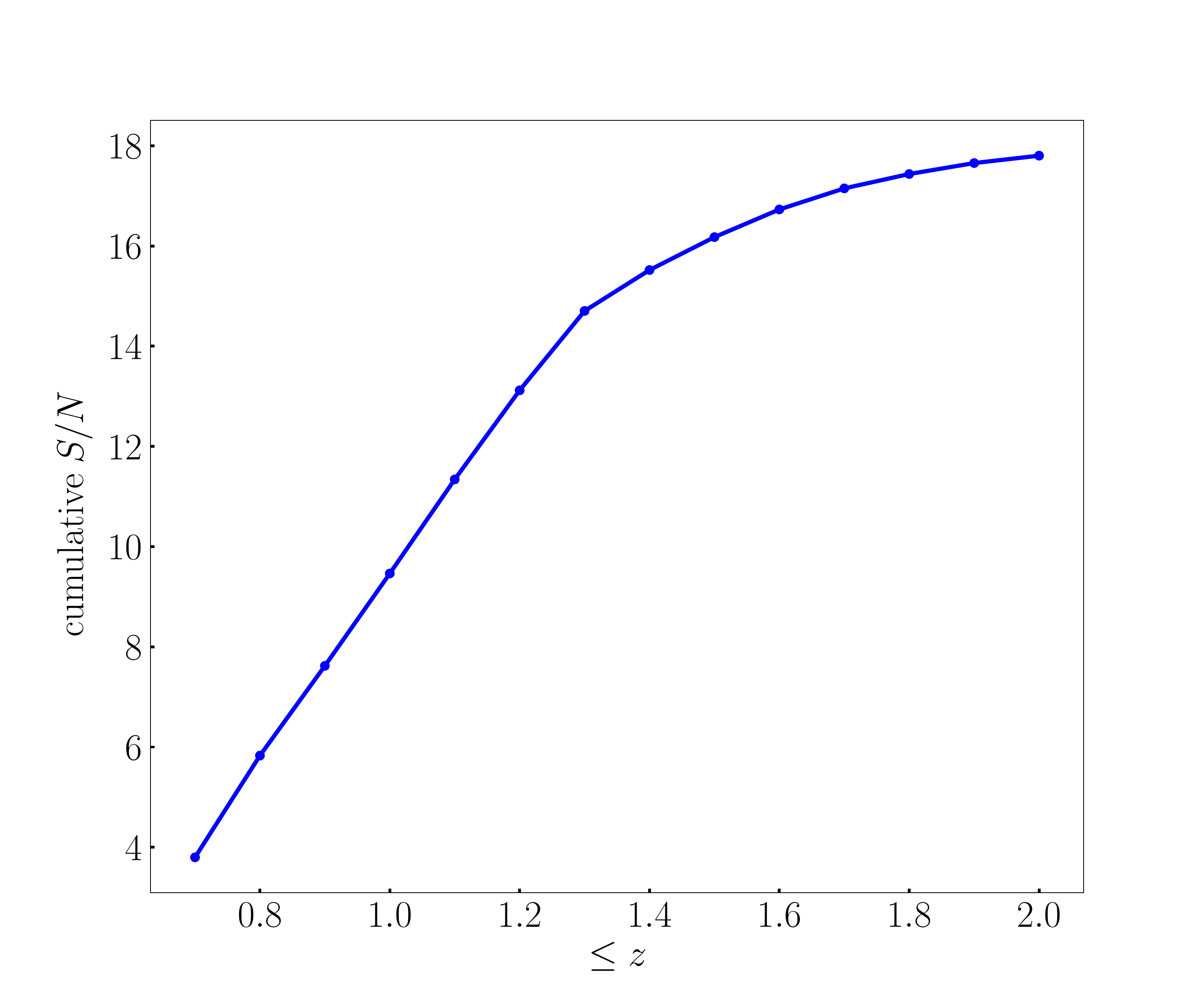}
\caption{Relativistic SNR per $z$-bin  ({\em left}) and cumulative ({\em right}) for a Stage IV $H\alpha$ survey.} \label{fig4}
\end{figure}
\begin{figure}[!ht]
\centering
\includegraphics[width=7.5cm]{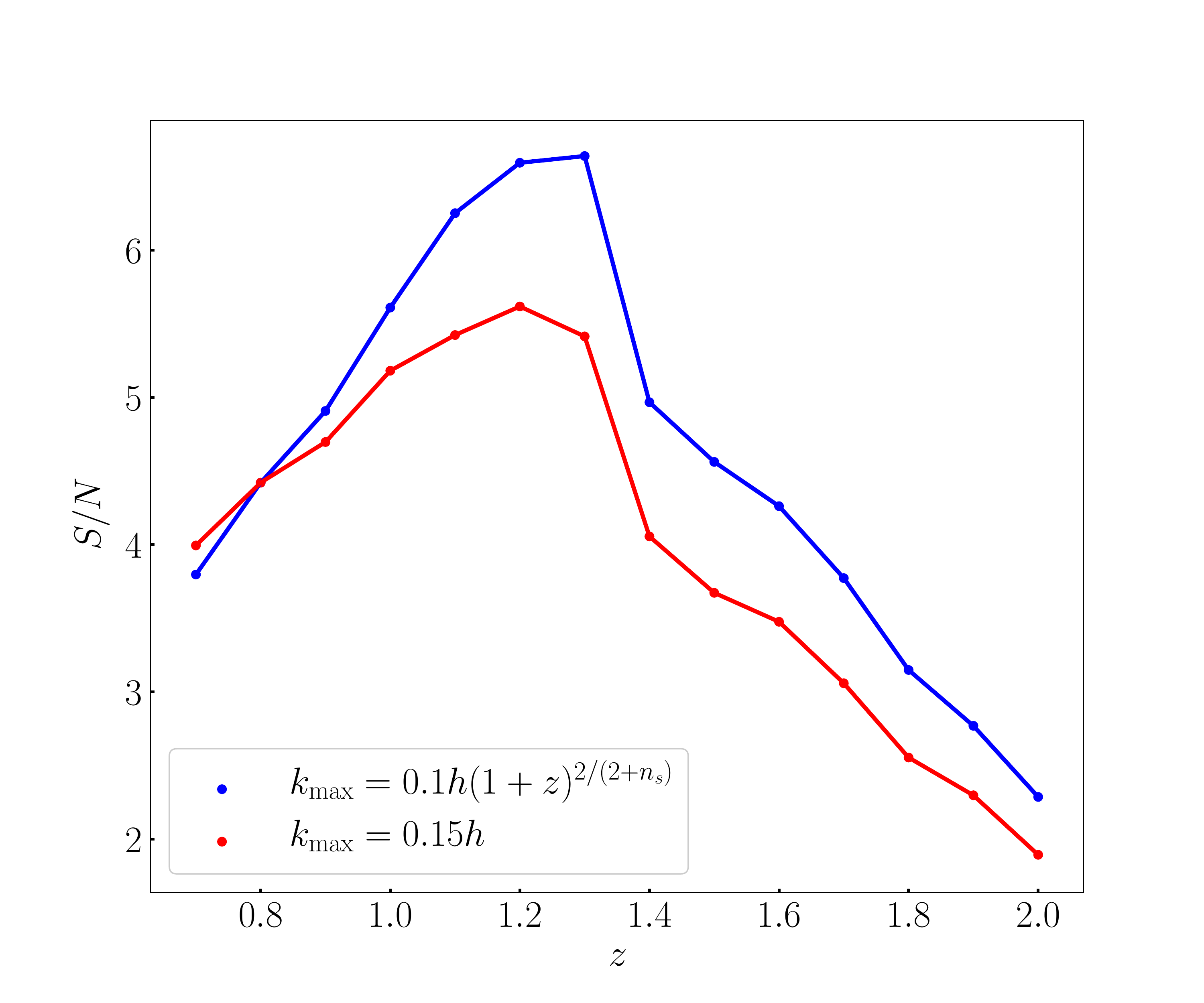}
\includegraphics[width=7.5cm]{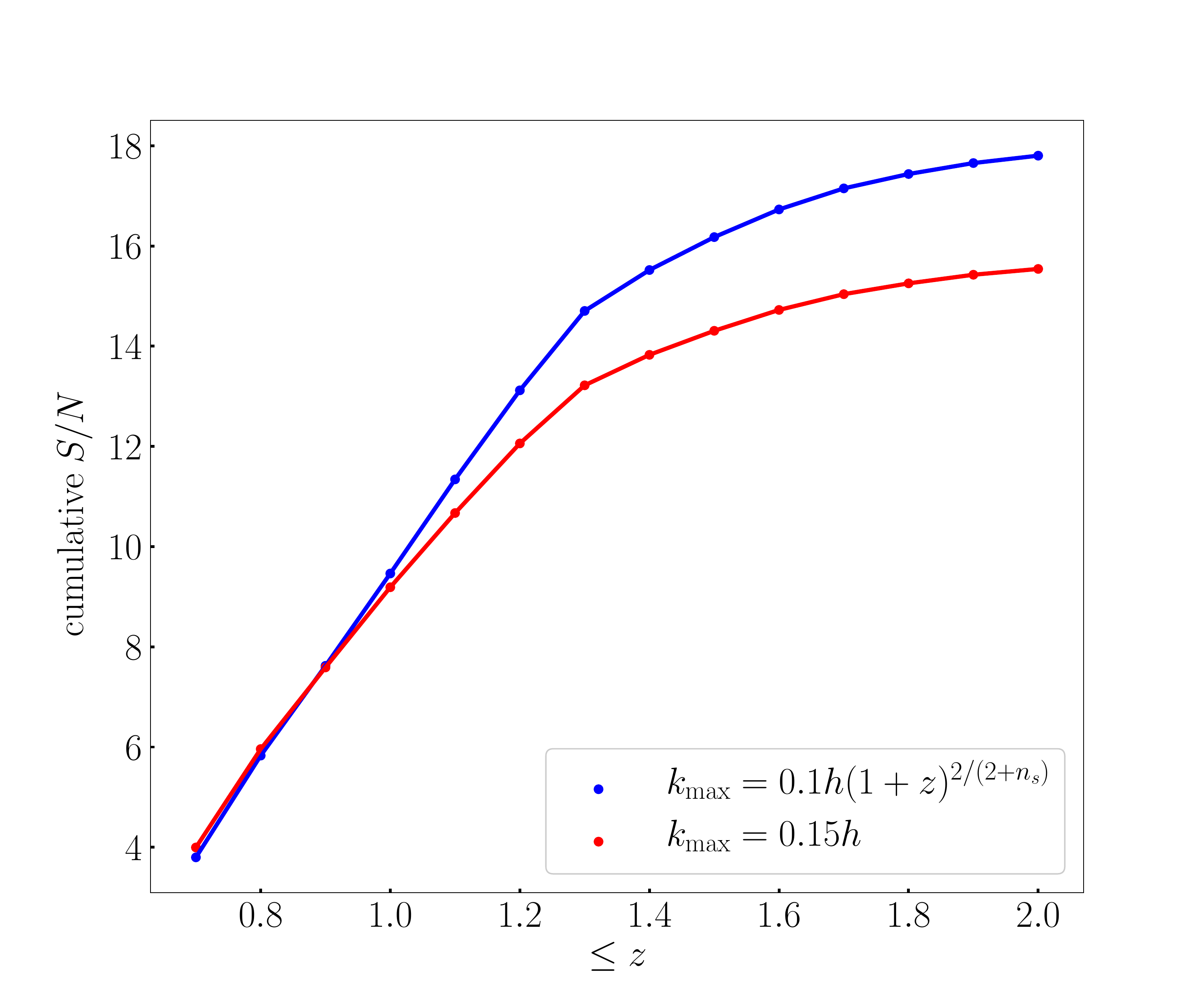} 
\caption{Effect of changing $k_{\rm max}$ on SNR per bin ({\em left}) and cumulative SNR ({\em right}).} \label{kmsnr}
\end{figure} 
\item  
Changes in $b_e(z),\Q(z)$: In the Appendix (Fig.~\ref{fig1x}) we  illustrate the significant impact on cumulative SNR of changing $b_e,\Q$. We use a range of constant choices for $b_e,\Q$ -- which are not physically motivated. This shows the importance of modelling $b_e,\Q$ self-consistently from the same  luminosity function that produces the number density, as we have done. 
\end{itemize}

{The sensitivity of the relativistic SNR to  $k_{\rm max}$ reflects the importance of the coupling of the relativistic signal to Newtonian terms on short scales. How sensitive is the SNR to the signal on the largest scales? We can answer this by increasing $k_{\rm min}$ from its fiducial value $k_{\rm f}$, which is the maximal observable scale. The result is that there is only a small reduction when $k_{\rm min}/k_{\rm f}$ is increased by a factor up to 5, as shown in Fig.~\ref{kmin}. 
{Even with $k_{\rm min}=10 k_{\rm f}$, the total SNR is $\sim 10$.} 
This means that the relativistic SNR does not depend critically on accessing the largest possible scales.}
\begin{figure}[! h]
\centering
\includegraphics[width=8.0cm]{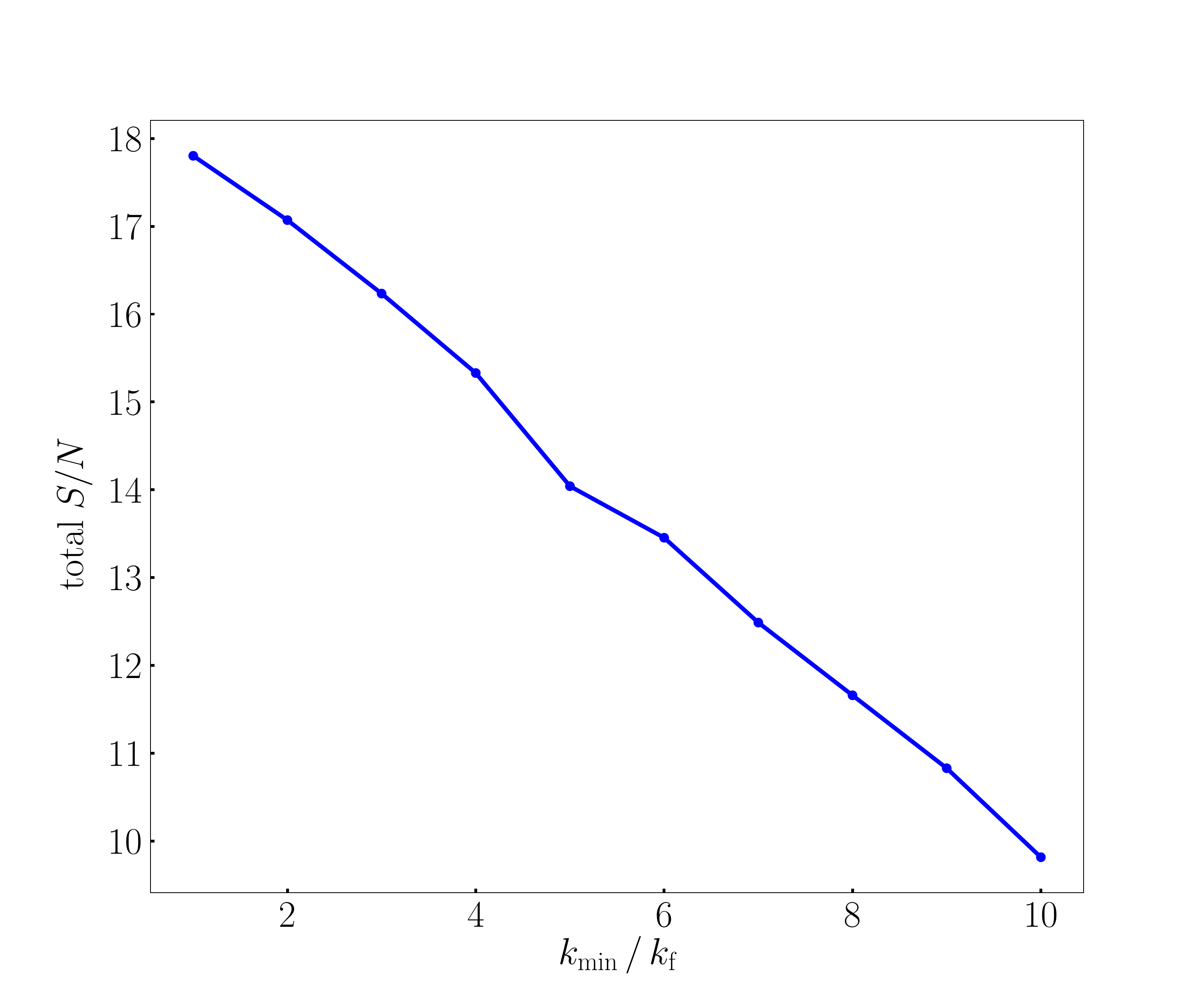}
\caption{{Effect of changing $k_{\rm min}$ on  total relativistic SNR.} 
} \label{kmin}
\end{figure}

It is also interesting to investigate how important for the SNR is the second-order relativistic contribution 
in the bispectrum, i.e. from terms of the form
\be\label{2oc}
\mathcal{K}^{(1)}_{\mathrm{N}}(\bm{k}_{1})\mathcal{K}^{(1)}_{\mathrm{N}}(\bm{k}_{2})\mathcal{K}^{(2)}_{\mathrm{D}}(\bm{k}_{1},\bm{k}_{2},\bm{k}_{3})\,,
\ee
 in \eqref{e21},  compared to the first-order contribution, i.e. from terms of the form
 \be \label{1oc}
 \Big[\mathcal{K}^{(1)}_{\mathrm{N}}(\bm{k}_{1})\mathcal{K}^{(1)}_{\mathrm{D}}(\bm{k}_{2}) + \mathcal{K}^{(1)}_{\mathrm{D}}(\bm{k}_{1})\mathcal{K}^{(1)}_{\mathrm{N}}(\bm{k}_{2})\Big]\mathcal{K}^{(2)}_{\mathrm{N}}(\bm{k}_{1},\bm{k}_{2},\bm{k}_{3}) \,.
 \ee
 It is conceivable that the first-order Doppler-type contribution in \eqref{1oc} to $B_g$, which couples to first- and second-order Newtonian terms, dominates the SNR. However, we find that the first- and  second-order relativistic parts of the bispectrum make comparable contributions to the SNR -- see Fig.~\ref{fig4z}. We deduce that the second-order relativistic contribution in \eqref{2oc} cannot be neglected. Furthermore, this means that it must be accurately modelled, as we  have done. 
\begin{figure}[! ht]
\centering
\includegraphics[width=7.5cm]{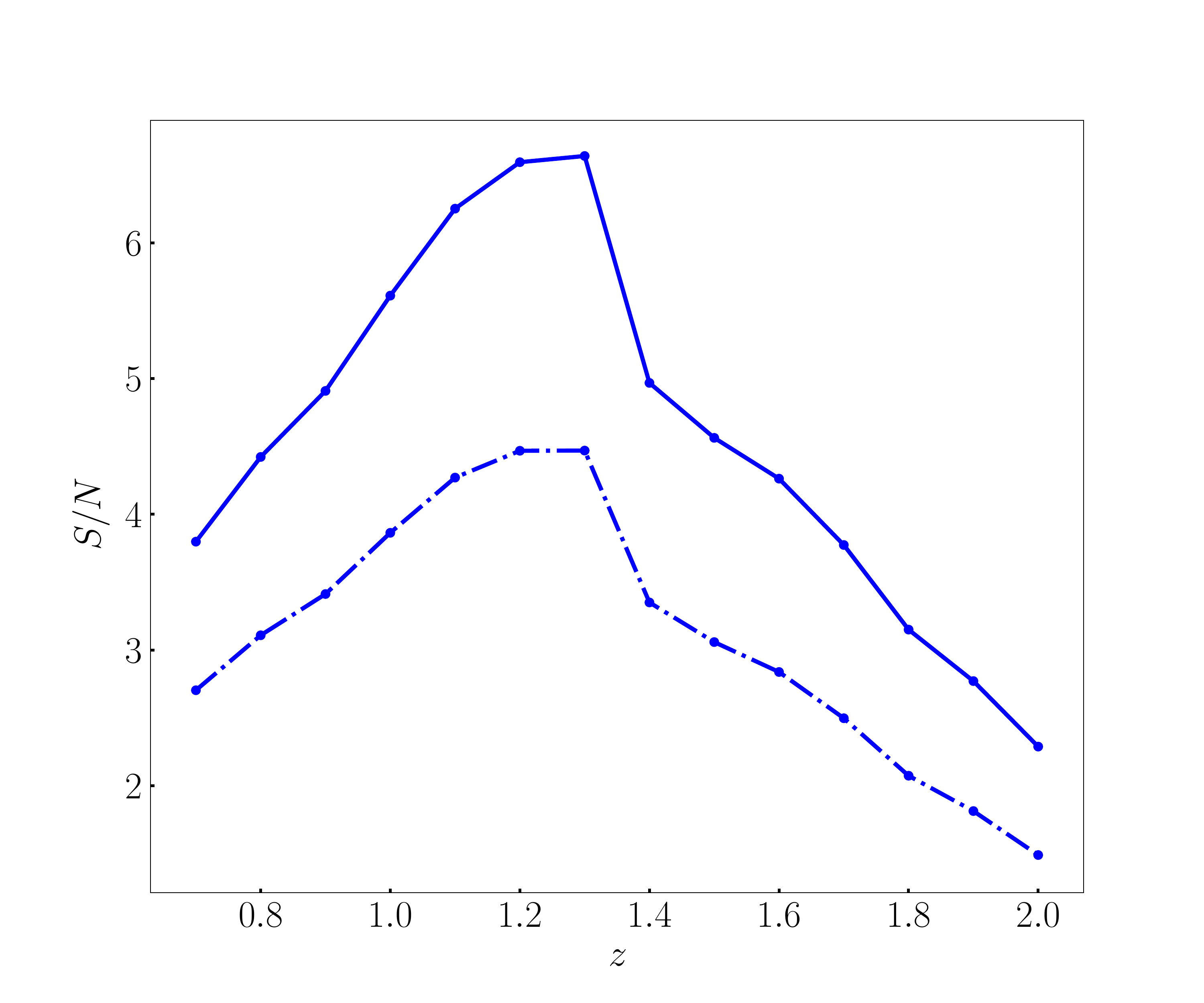}
\includegraphics[width=7.5cm]{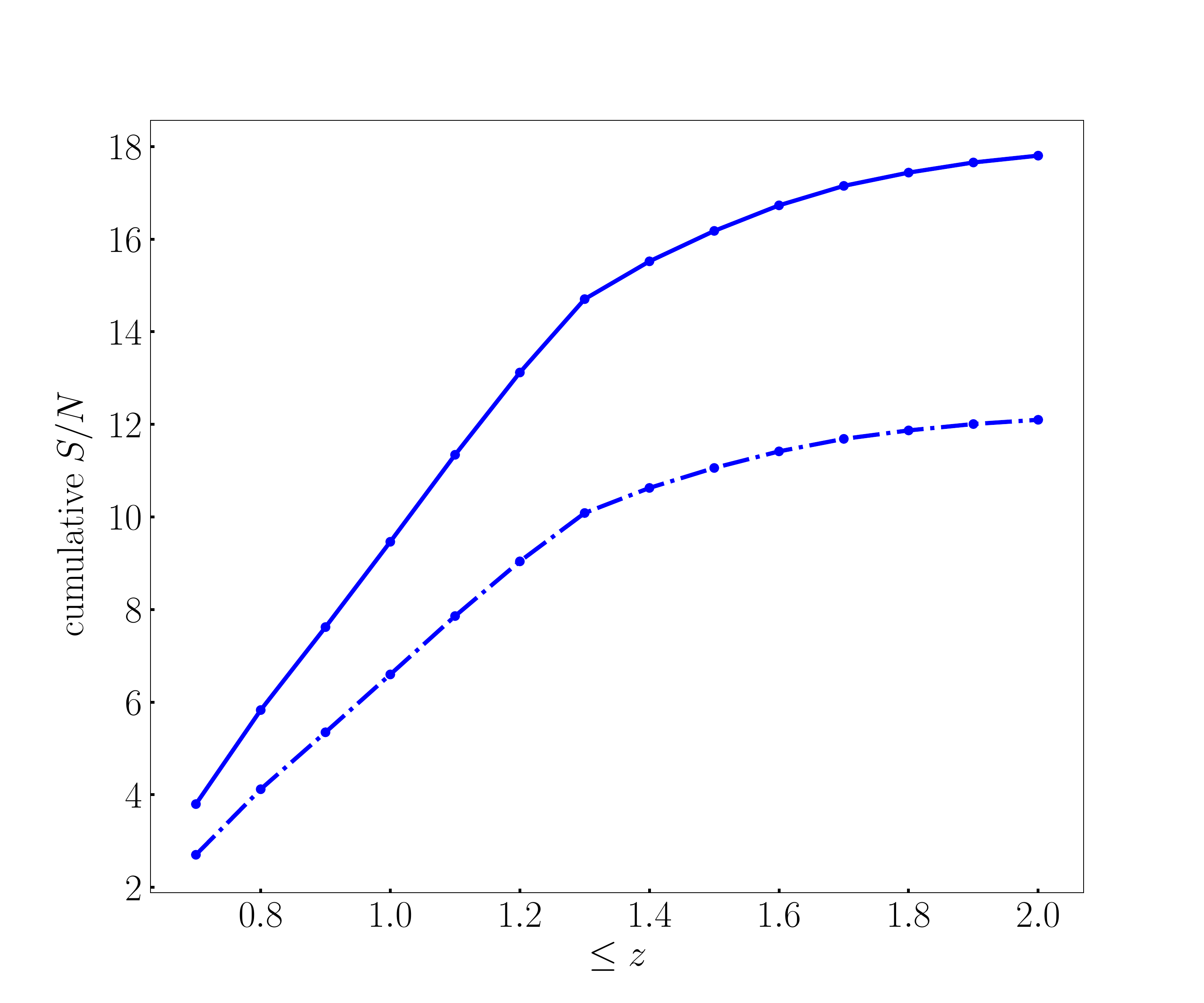}
\caption{As in Fig.~\ref{fig4}, but showing the effect of omitting the second-order relativistic contribution \eqref{2oc} to the bispectrum (dot-dashed curves).} \label{fig4z}
\end{figure}

Recently \cite{Jeong:2019igb} estimated the  SNR for the leading relativistic part of the bispectrum.  There are significant differences in their analysis compared to ours. In particular, they neglect most of the terms in $\delta_{g \rm D}^{(2)}$ [see our \eqref{dg2}] which defines ${\cal K}^{(2)}_{\rm D}$ (see the Appendix for further details). {In addition they do not use self-consistent models for $b_{e}$ and $\mathcal{Q}$. These two differences could account for their conclusion that the relativistic signal is not detectable, in contrast to our result.} 

{An interesting feature of the relativistic signal is that there is a significant contribution to the SNR from  flattened triangle shapes. This is consistent with the results of  \cite{Clarkson:2018dwn} for the dipole that is generated by the imaginary part of the bispectrum.}

~\\
\subsection{Including cosmological parameters}
~\\
\begin{figure}[! h]
\centering
\includegraphics[width=8.0cm]{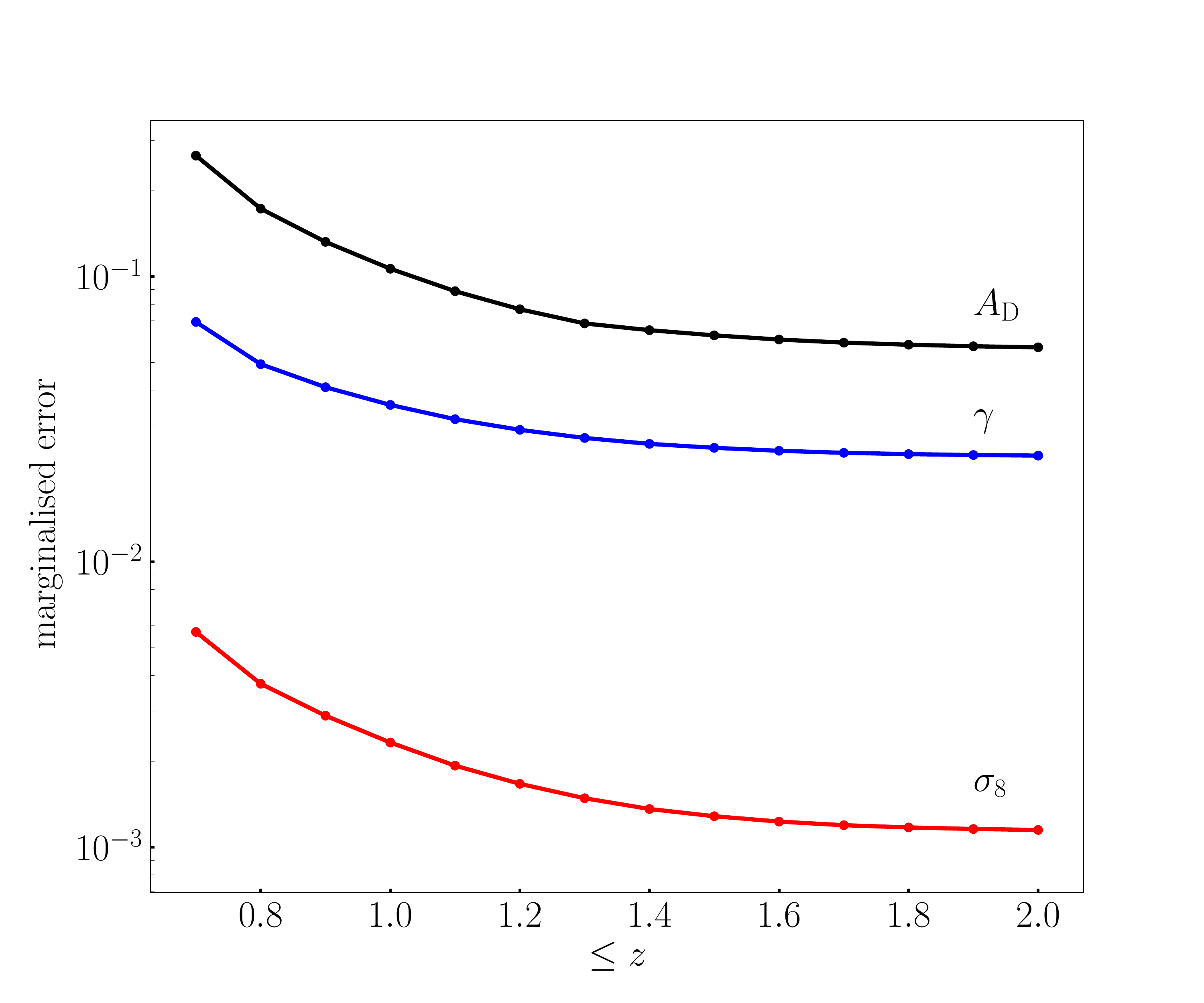}
\caption{Marginal errors (cumulative) on the relativistic contribution, growth index and clustering amplitude, using the full bispectrum.} \label{sig}
\end{figure}
{A full treatment of cosmological constraints would marginalise over the standard cosmological parameters, together with the Alcock-Paczynski parameters and the clustering bias parameters. The constraints obtained would depend almost entirely on the Newtonian galaxy power spectrum and bispectrum (as analysed in \cite{Yankelevich:2018uaz}), given that the relativistic contribution to the power spectrum is below leading order, while in the bispectrum the relativistic SNR is  an order of magnitude smaller than the Newtonian SNR. 

Our focus here is instead on the detectability of the relativistic signal in the bispectrum, {assuming that $P_{g\rm N}$ and $B_{g\rm  N}$ have been used to constrain the standard parameters.}
We now investigate the effect on this detectability when we include the parameters directly related to redshift-space effects, i.e.,  the growth index $\gamma=\ln f/ \ln\Omega_m$ (with fiducial value 0.545), and the clustering amplitude $\sigma_8$ (with Planck 2018 fiducial).} 
{For forecasts we use the theoretical values of $b_e, \Q$. In a galaxy survey, they would be measured directly from the observed luminosity function, and their measurement uncertainties would need to be marginalised over.}

For the relativistic part of the bispectrum, we introduce a parameter  $A_{\rm D}$, with fiducial value 1: 
\be
B_g=B_{g\rm N}+A_{\rm D}\,B_{g\rm D} \,.
\ee
Then the Fisher matrix for the parameters  $\vartheta_\alpha=(A_{\rm D},\gamma,\sigma_8)$ is
\begin{equation}
F_{\alpha\beta} = 
\sum_{z,k_a,\,\mu_{1},\,\varphi}\,{1\over {\rm Var} [{B_{g}}]}
\,{\partial B_{g}\over \partial \vartheta_{(\alpha}}\,{\partial B^*_{g}\over \partial \vartheta_{\beta)}}\,,\label{fm}
\end{equation} 
where the round brackets denote symmetrisation. The cumulative marginal errors $\sigma_\alpha = [(F^{-1})_{\alpha\alpha}]^{1/2}$  are shown in Fig.~\ref{sig}. The fact that $\sigma_{A_{\rm D}}\lesssim 0.1$  means that the relativistic effects remain detectable when the two additional cosmological parameters are marginalised over.

\clearpage
\section{{CONCLUSIONS}}

As shown by \cite{Clarkson:2018dwn}, the tree-level galaxy bispectrum in Fourier space has an imaginary part which is a unique signal of the leading-order relativistic corrections in redshift space. These corrections arise from Doppler and other line-of-sight effects on the past lightcone [see \eqref{dg1}, \eqref{dg2}]. In the galaxy bispectrum, the corrections scale as ${\rm i}\,(\cH/k)P^2$, where $P$ is the linear matter power spectrum [see \eqref{e21}--\eqref{e24}]. By contrast,  at leading order in the galaxy power spectrum, the relativistic correction is real and scales as $(\cH/k)^2P$ -- i.e., it is suppressed by a further factor of  $\cH/k$. 
Only the cross-power spectrum of two different tracers produces an imaginary contribution that scales as ${\rm i}\,(\cH/k)P$ \cite{McDonald:2009ud}.\footnote{See also \cite{Bonvin:2013ogt, Bonvin:2015kuc, Gaztanaga:2015jrs, Irsic:2015nla, Hall:2016bmm,Lepori:2017twd,Bonvin:2018ckp, Lepori:2019cqp} for the corresponding effect in the two-point correlation function, and see \cite{Okoli:2016vmd} for an imaginary {short-scale} contribution from neutrino drag on haloes.}

For a single tracer, the $(\cH/k)^{2}$ relativistic signal in the galaxy {\em power} spectrum is not detectable, even for a cosmic-variance limited survey \cite{Alonso:2015uua}.
The galaxy bispectrum of a single tracer, with its ${\rm i}\,(\cH/k)$ relativistic contribution, improves the chances of detectability. In addition, the relativistic contribution in the bispectrum couples to short-scale Newtonian terms -- which means that the signal is not confined to very large scales, unlike the case of the power spectrum. We confirmed the expectations of detectability by showing that the signal to noise on the imaginary relativistic part is ${\cal O}(10)$ for a Stage IV $H\alpha$ spectroscopic survey similar to Euclid [see Fig. \ref{fig4}].  We checked that detectability is not compromised by including the uncertainties on two cosmological growth parameters, $\sigma_8$ and $\gamma$ [Fig. \ref{sig}], assuming that other cosmological  and nuisance parameters are determined by the Newtonian power spectrum and bispectrum.

The relativistic SNR depends on the $k_{\rm max}(z)$ assumed, because of the coupling of relativistic effects to short-scale Newtonian terms [Fig.~\ref{kmsnr}], and  we made a conservative choice \eqref{kmax}, which includes a redshift dependence to reflect the weakening of nonlinearity at higher $z$. 
Accurate modelling of nonlinear effects would allow us to increase the SNR  
 -- this is not at all specific to the relativistic signal,  but is required for the standard analysis of RSD.

The relativistic SNR also relies  on the largest available scales, but very little signal is lost if $k_{\rm min}/k_{\rm f}$ is increased by a factor up  to 5, and even a factor of 10 increase leaves a detectable SNR  [Fig. \ref{kmin}].

By contrast, the SNR depends strongly on accurate modelling of the second-order part of the relativistic correction  [Fig.~\ref{fig4z}]. This includes both the theoretical form \eqref{dg2}, and the
 two astrophysical parameters that do not appear in the Newtonian approximation of the galaxy bispectrum: the evolution bias $b_e$ (measuring the deviation from comoving number conservation) and the magnification bias $\Q$ (which is brought into play by a Doppler correction to standard lensing magnification).  A key feature of our analysis is a physically self-consistent derivation of these quantities from the luminosity function [Fig.~\ref{fig2} and \eqref{q}, \eqref{be}].
We showed that the SNR is very sensitive to these parameters [Fig. \ref{fig1x}], which underlines the need for accurate physical modelling. 

We assumed a Gaussian covariance in our computations, but we used the approximation  of \cite{Chan:2016ehg} to include non-Gaussian corrections. 

Further work should include the window function which we have  neglected.
The imaginary part of the galaxy bispectrum generates a dipole, as shown in \cite{Clarkson:2018dwn}. This suggests a multipole analysis that uses the relativistic dipole in addition to the monopole and quadrupole, {which are unaffected by relativistic effects at leading order.} The
window function can also have an imaginary part \cite{Beutler:2018vpe}, which will need to be corrected for. The dipole from the imaginary part of the bispectrum vanishes in equilateral configurations \cite{Clarkson:2018dwn}, which may help to disentangle the relativistic dipole from that of the window function.

{Our analysis, in common with other works on the Fourier bispectrum, implicitly uses the plane-parallel approximation, since the line-of-sight direction ${\bm{n}}$ is fixed. At the cost of significant complexity, the approximation can be avoided, for example by using a Fourier-Bessel analysis of bispectrum multipoles \cite{Castorina:2017inr}. Further work is needed to address this, but we note that errors from the approximation are mitigated in high redshift surveys such as the one considered here.}

Finally, further work also needs to include the effects of lensing magnification, which are excluded in the standard Fourier analysis, but have been included in the galaxy angular bispectrum \cite{Kehagias:2015tda,DiDio:2015bua, DiDio:2016gpd, DiDio:2018unb} and in a spherical Bessel analysis \cite{Bertacca:2017dzm}.  

\vfill
\noindent {\bf Acknowledgements}\\
{\small {We thank Dionysis Karagiannis and Shun Saito for very helpful comments.}
RM and SJ are supported by  the South African Radio Astronomy Observatory (SARAO) and the National Research Foundation (Grant No. 75415). 
 RM and OU are supported by the UK Science \& Technology Facilities Council (STFC) Consolidated Grants ST/N000668/1 and ST/K0090X/1. 
CC is supported by STFC Consolidated Grant ST/P000592/1.
SC acknowledges support from the `Departments of Excellence 2018-2022' Grant (L.\ 232/2016) awarded by the Italian Ministry of Education, University and Research (\textsc{miur}). SC is funded by \textsc{miur} through the Rita Levi Montalcini project `\textsc{prometheus} -- Probing and Relating Observables with Multi-wavelength Experiments To Help Enlightening the Universe's Structure'.} 

\clearpage
\appendix

\section{APPENDIX}

\subsection*{Newtonian kernels in \eqref{k2n}}
\begin{eqnarray}
F_{2}(\bm{k}_{1}, \bm{k}_{2}) &=& \frac{10}{7} + {\hat{\bm{k}}_{1} \cdot \hat{\bm{k}}_2}\bigg(\frac{k_{1}}{k_{2}} + \frac{k_{2}}{k_{1}}\bigg) + \frac{4}{7}\big({\hat{\bm{k}}_{1} \cdot \hat{\bm{k}}_2}\big)^{2}\,,
\label{e18} \\
G_{2}(\bm{k}_{1}, \bm{k}_{2}) &=& \frac{6}{7} + {\hat{\bm{k}}_{1} \cdot \hat{\bm{k}}_2}\bigg(\frac{k_{1}}{k_{2}} + \frac{k_{2}}{k_{1}}\bigg) + \frac{8}{7}\big({\hat{\bm{k}}_{1} \cdot \hat{\bm{k}}_2}\big)^{2} \label{e19}\,,\\
Z_2(\bm{k}_1,\bm{k}_2) &=&
 f{\mu_1\mu_2 \over k_1k_2}\big( \mu_1k_1+\mu_2k_2\big)^2 
+{b_{1}\over k_1k_2}\Big[ \big(\mu_1^2+\mu_2^2 \big)k_1k_2+\mu_1\mu_2\big(k_1^2+k_2^2 \big) \Big]\,, \label{e16x} \\  
S_{2}(\bm{k}_{1}, \bm{k}_{2}) &=& \big({\hat{\bm{k}}_{1} \cdot \hat{\bm{k}}_2}\big)^{2} - \frac{1}{3}\,. \label{e20}
\end{eqnarray}

\subsection*{Fitting formulas for Fig.~\ref{fig1} curves}
\bea
b_1(z) &=& 0.9+0.4 z\,, \quad b_2(z) = -0.741 - 0.125\,z + 0.123\,z^{2} + 0.00637\,z^{3}\,, \label{e9}  \\
 b_{s^2}(z) &=& 0.0409 - 0.199\,z - 0.0166\,z^{2} + 0.00268\,z^{3} \,, \label{e11} \\
 V(z) &=& 8.85\, z^{1.65}\exp\big({-0.777\,z}\big)~ h^{-3}\,\mathrm{Gpc}^{3}\,, \label{e11_1} \\
n_{g}(z) &=& 0.0193\, z^{-0.0282}\exp\big({-2.81\,z}\big) ~h^{3}\,\mathrm{Mpc}^{-3}\,,
 \label{e11_2}\\
\sigma(z)& =& \big(5.29 - 0.249\,z - 0.720\,z^{2} + 0.187\,z^{3}\big)~h^{-1}\,\mathrm{Mpc}\,. \label{e14a}
\eea

\subsection*{{Number of orientation bins}}

Figure~\ref{fig4x} 
shows the effect on the 
relativistic 
total SNR of changing the number of orientation bins, $n_{\mu_1}=N_{\mu_1}/\Delta{\mu_1}=2 /\Delta{\mu_1}$  and $n_\varphi=N_\varphi/\Delta\varphi =2\pi/\Delta\varphi$. It is apparent 
that reducing the number of bins increases the cumulative SNR. The cumulative SNR 
converges towards a minimum for 
$n_{\mu_{1}}\,, n_{\varphi} > 40$.
We choose $n_{\mu_{1}}= n_{\varphi}=50$, which is equivalent to \eqref{ori}.
\begin{figure}[! ht]
\centering
\includegraphics[width=8.0cm]{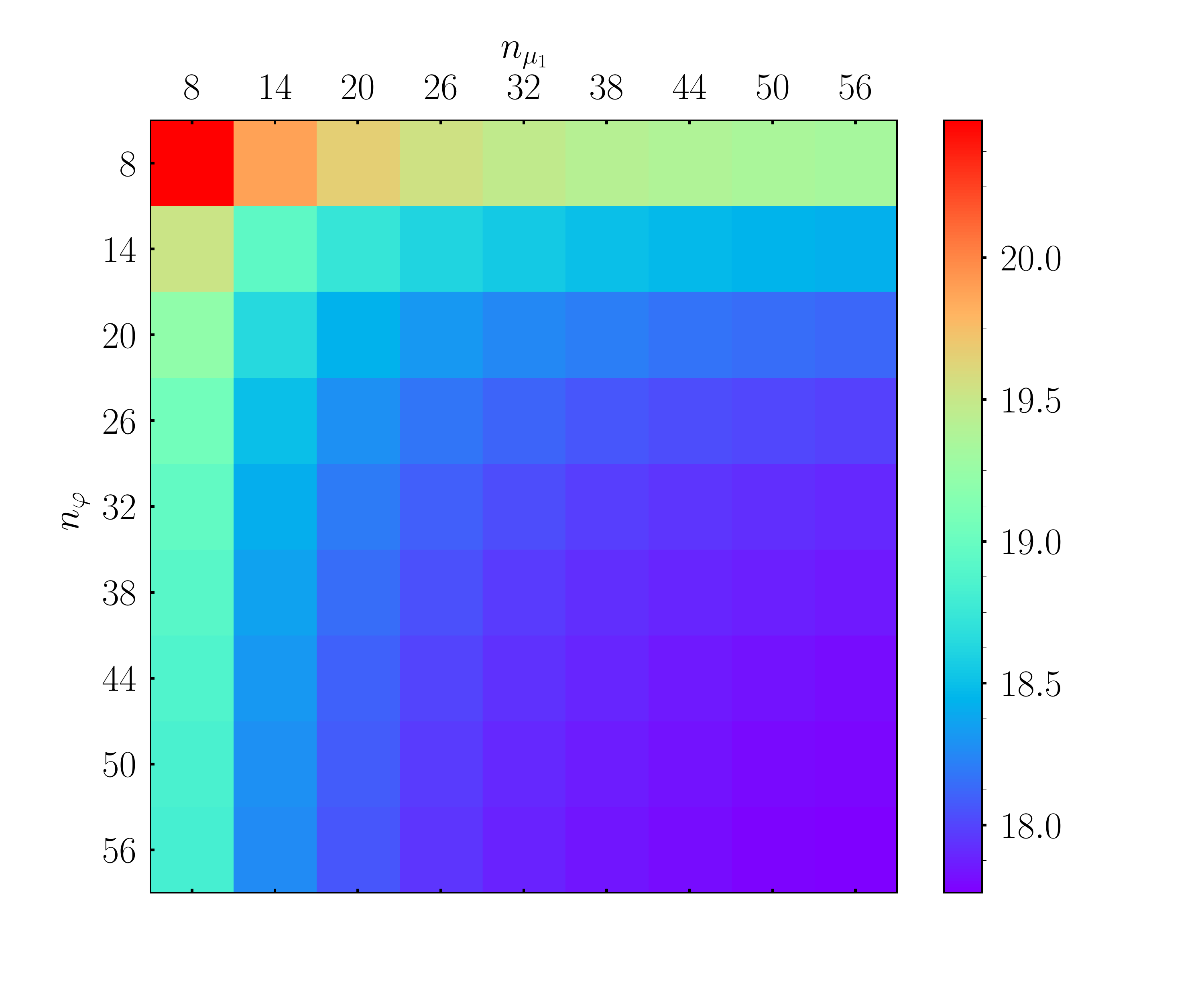} 
\caption{Effect on 
{total} relativistic SNR of changing number of  $\varphi$ and $\mu_1$ bins. 
} \label{fig4x}
\end{figure}

\subsection*{{Effect of changing magnification and evolution biases}}

The effect on the relativistic SNR of changes in magnification bias and in evolution bias is illustrated in Fig.~\ref{fig1x}.

\begin{figure}[! h]
\centering
\includegraphics[width=7.5cm]{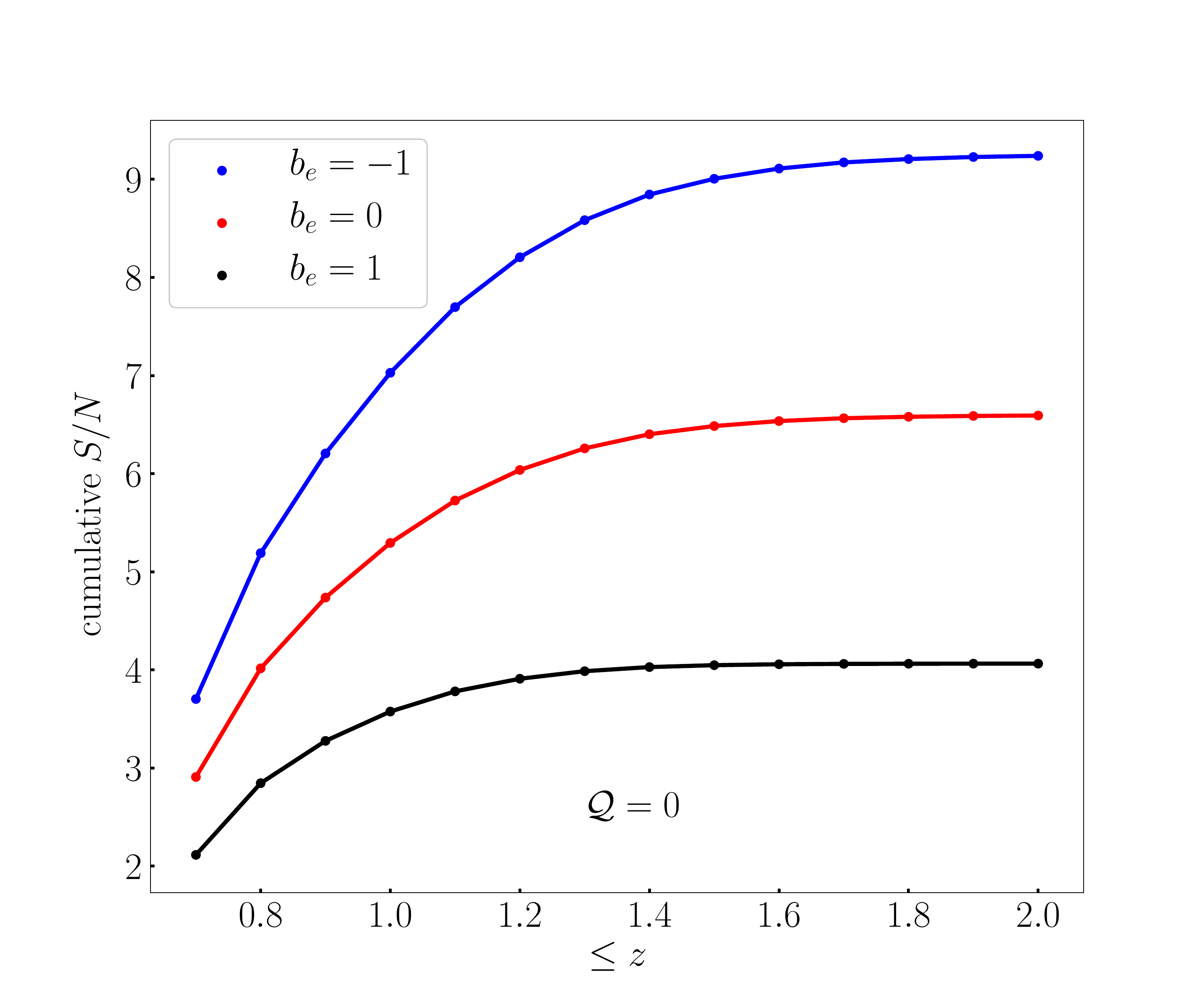} 
\includegraphics[width=7.5cm]{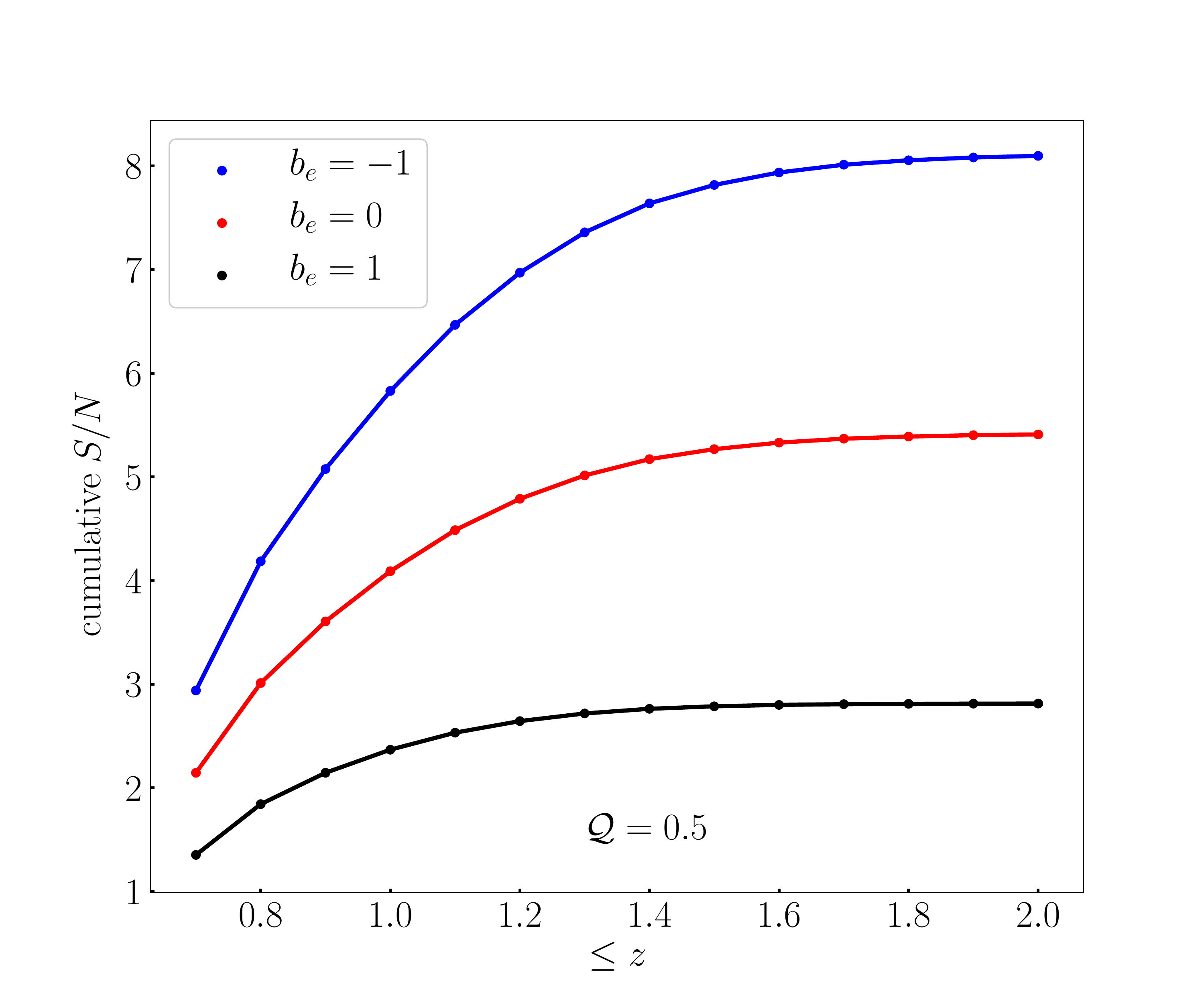} \\
\includegraphics[width=7.5cm]{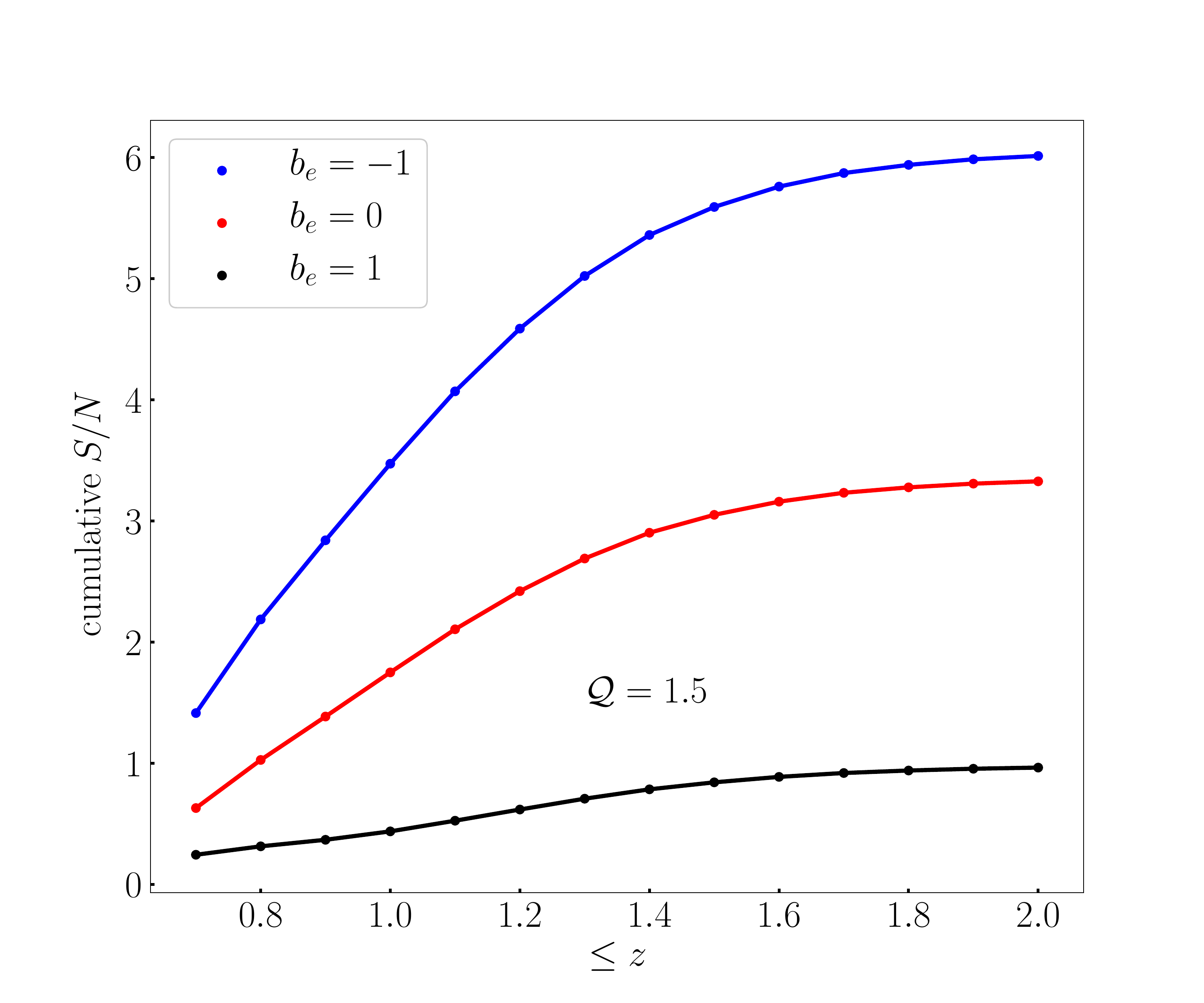}
\includegraphics[width=7.5cm]{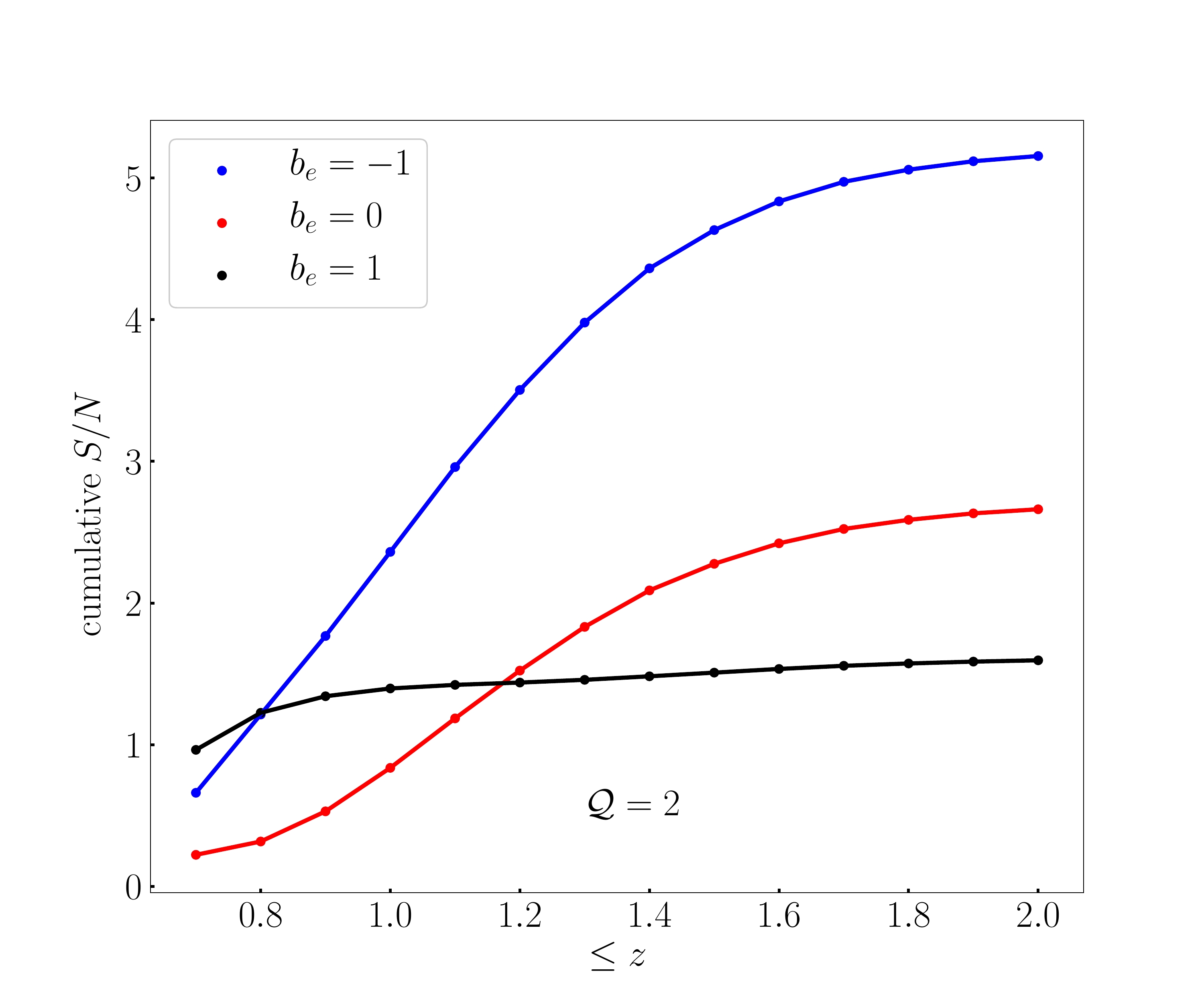} 
\caption{Effect of changing ${\cal Q}$ and $b_e$ on relativistic cumulative SNR.} \label{fig1x}
\end{figure}

\subsection*{{Comparison with \cite{Jeong:2019igb}}}

In \cite{Jeong:2019igb},  
a significant number of terms is neglected in the relativistic second-order  galaxy number count contrast, $\delta_{g\rm D}^{(2)}$, given by our \eqref{dg2}. (Note that our  \eqref{dg2}, derived in \cite{Clarkson:2018dwn}, was independently confirmed by \cite{DiDio:2018zmk}). They have the first term, $A\, \bm{v}^{(2)}\!\!\cdot\bm{n}$, on the right of \eqref{dg2}.  In the second term, $2{C}(\bm{v}\cdot\bm{n})\,\delta$, they do not have the correct form of the coefficient $C$ -- they include only the first part, $b_1A$, of $C$ [see the right-hand side of \eqref{e26}]. All terms after the second term in \eqref{dg2} are omitted by \cite{Jeong:2019igb}. Note that none of the omitted terms is suppressed by a higher power of $k^{-1}$;  they all have the same scaling, i.e., $\propto (\cH/k)\,{(\delta)^2}$. In detail, they omit the following
terms:
\bea
\delta_{g\rm D}^{(2)}({\rm us})- \delta_{g\rm D}^{(2)}(\mbox{{\cite{Jeong:2019igb}}}) &=& 2\left[b_{1}f + \frac{b_1'}{\cH} 
+ 2\bigg(1-\frac{1}{{r} \cH}\bigg){\frac{\partial b_1}{\partial \ln{L}}\bigg|_{\rm c}} \right](\bm{v}\cdot\bm{n})\,\delta 
\\\nonumber
&&{}
+{2\over\cH}\left(4-2A-\frac{3}{2}\Omega_{m} \right)(\bm{v}\cdot\bm{n})\,\partial_r(\bm{v}\cdot\bm{n})\\
\nonumber
&&{}
+{2\over\cH^2}\big[(\bm{v}\cdot\bm{n}) \,\partial_r^2\Phi-\Phi\, \partial_r^2 (\bm{v}\cdot\bm{n}) \big]
 -{2\over \cH}\,\partial_r (\bm{v}\cdot\bm{v})+2{b_1\over\cH}\,\Phi\, \partial_r\delta \,. 
\eea 


\clearpage
\bibliographystyle{JHEP}
\bibliography{reference_library}

\end{document}